\RequirePackage{rotating}
\documentclass[sigconf]{acmart}

\usepackage{amsmath}
\usepackage{amsmath}
\usepackage[T1]{fontenc}
\usepackage{graphicx}
\usepackage{textcomp}
\usepackage{tikz}
\usepackage{hyperref}
\usepackage{textcomp}
\usepackage{color}
\usepackage[utf8]{inputenc}
\usepackage{fancyvrb}
\usepackage{xcolor}
\usepackage{colortbl}
\usepackage{hhline}
\usepackage{lscape}
\usepackage{upquote}
\usepackage{array}
\usepackage{rotating}
\usepackage{stmaryrd}
\usepackage{pifont}
\usepackage{multirow}
\usepackage{wrapfig}
\usepackage{verbatim}
\usepackage{colortbl}
\usepackage{rotating}
\usepackage{algorithm}
\usepackage{algorithmic}
\usepackage{xspace}
\usepackage{flushend}
\usepackage{balance}
\usepackage{url}
\usepackage[export]{adjustbox}
\usepackage{listings}
\usepackage{pgfplotstable}
\pgfplotsset{compat=1.8}
\usepgfplotslibrary{statistics}
\usepackage{pgfplots}
\usepgfplotslibrary{colorbrewer}
\pgfplotsset{compat = 1.15, cycle list/Set1-8} 
\usetikzlibrary{colorbrewer} 
\usepackage{pgfplotstable}
\usepackage{filecontents}
\usepackage{framed}
\usepackage[strict]{changepage}    
\usepackage{pgf-pie}

\newcolumntype{L}[1]{>{\raggedright\arraybackslash}m{#1}}
\newcolumntype{C}[1]{>{\centering\arraybackslash}m{#1}}
\newcolumntype{R}[1]{>{\raggedleft\arraybackslash}m{#1}}

\fvset{commandchars=\\\{\}}

\definecolor{result}{rgb}{235, 235, 235}
\definecolor{DarkGreen}{rgb}{0,0.3,0}
\definecolor{DarkBlue}{rgb}{0,0,0.7}
\lstdefinelanguage{SMV}[]{}{
  commentstyle=\color{DarkGreen}\itshape,
  keywordstyle=\color{blue}\bfseries,
  morekeywords=[1]{VAR,INIT,LTLSPEC,TRANS,GAR,ASM,TRUE,X,G,F,CTLSPEC,AG,EX,MODULE,next},
  morekeywords=[1]{occurs,between,and,GF,leads,to,After,have,at,most,two,until,Globally,S},
  morecomment=[l]{--}
}
\lstdefinelanguage{Alloy}[]{Java}{
commentstyle=\color{DarkGreen}\itshape,
  morekeywords={module,abstract,sig,->,fact,pred,fun,run,for,iff,implies,
  not,no,one,all,some,lone,\#,set,in,and,or,but,exactly,none,univ,Int,assert,check},
	  otherkeywords = {<,<->,->, &, |, =, !=, !,<:,~},
	  morekeywords = {<,<->,->, &, |, =, !=, !,<:,~},
	  keywordstyle={\color{blue}}
  }

\definecolor{Green}{rgb}{0.0, 0.56, 0.0}
\definecolor{Gray}{gray}{0.85}
\newcommand{\Blue}[1]{\textcolor{blue}{\textbf{#1}}}
\newcommand{\Green}[1]{\textcolor{Green}{\textbf{#1}}}
\newcommand{\Red}[1]{\textcolor{red}{\textbf{#1}}}

\newcommand{\CodeIn}[1]{\begin{small}\texttt{#1}\end{small}}
\newcommand{\Comment}[1]{}
\newcommand{\NoType}[1]{} % we don't use goal type in expression generation
\newcommand{\Space}[1]{}

\newcommand{\DefMacro}[2]{\expandafter\newcommand\csname rmk-#1\endcsname{#2}}
\newcommand{\UseMacro}[1]{\csname rmk-#1\endcsname}

\newcommand{\ACaret}{\raisebox{-0.6ex}{\textasciicircum}}
\newcommand{\AStar}{*}

\newcommand\diff\setminus

\definecolor{formalshade}{rgb}{0.95,0.95,1}
\definecolor{mygray}{rgb}{0.80,0.80,1}
\definecolor{cadetgrey}{rgb}{0.57, 0.64, 0.69}
\definecolor{silver}{rgb}{0.75, 0.75, 0.75}
\definecolor{whitesmoke}{rgb}{0.96, 0.96, 0.96}

\newenvironment{impact}{%
  \MakeFramed{\advance\hsize-\width\FrameRestore}%
  \noindent\hspace{-4.55pt}% disable indenting first paragraph
  \begin{adjustwidth}{}{7pt}%
  \vspace{2pt}\vspace{2pt}%
}
{%
  \vspace{2pt}\end{adjustwidth}\endMakeFramed%
}
\pdfoutput=1
\begin{document}

\title{Right or Wrong -- Understanding How Novice Users Write Software Models}

\author{Ana Jovanovic}
\affiliation{%
  \institution{The University of Texas at Arlington}
  \country{Arlington, TX, USA}
}
\email{ana.jovanovic@mavs.uta.edu}

\author{Allison Sullivan}
\affiliation{%
  \institution{The University of Texas at Arlington}
  \country{Arlington, TX, USA}
}
\email{allison.sullivan@uta.edu}


\begin{abstract}
Writing declarative models has numerous benefits, ranging from automated reasoning and correction of design-level properties before systems are built, to automated testing and debugging of their implementations after they are built. Alloy is a declarative modeling language that is well-suited for verifying system designs. A key strength of Alloy is its scenario-finding toolset, the Analyzer, which allows users to explore all valid scenarios that adhere to the model’s constraints up to a user-provided scope. However, even with visualized scenarios, it is difficult to write correct Alloy models. To address this, a growing body of work explores different techniques for debugging Alloy models. In order to develop and evaluate these techniques in an effective manor, this paper presents an empirical study of over 97,000 models written by novice users trying to learn Alloy. We investigate how users write both correct and incorrect models in order to produce a comprehensive benchmark for future use as well as a series of observations to guide debugging and educational efforts for Alloy model development.
\end{abstract}

\maketitle

\section{Introduction}\label{sec:intro}
In today's society, we are becoming increasingly dependent on software systems. However, we also constantly witness the negative impacts of buggy software. One way to help develop better software systems is to leverage software models. When forming requirements, software models can be used to clearly communicate to all stakeholders both the desired system as well as the environment it will be deployed in. When creating designs and implementations, software models can help reason over how well the design and implementation choices satisfy the requirements. As such, software models can help detect flaws earlier in development and thus aid in the delivery of more reliable systems.

Alloy~\cite{JacksonAlloyBook2006} is a relational modeling language. A key strength of Alloy is the ability to develop models in the Analyzer, an automatic analysis engine based on off-the-shelf SAT solvers, which the Analyzer uses to generate scenarios that highlight how the modeled properties either hold or are refuted, as desired. The user is able to iterate over these scenarios one by one, inspecting them for correctness. Alloy has been used to verify software system designs~\cite{Zave15,BagheriETAL2018,WickersonETAL2017,ChongETAL2018},
and to perform various forms of analyses over the corresponding implementation, including deep static checking~\cite{JacksonVaziri00Bugs,TACOGaleottiETALTSE2013}, systematic testing~\cite{MarinovKhurshid01TestEra}, data structure repair~\cite{ZaeemKhurshidECOOP2010}, automated debugging~\cite{GopinathETALTACAS2011} and to generate security attacks~\cite{websecurity,Margrave,CheckMateMicro2019}. 

However, to gain the many benefits that come from utilizing software models, the model itself needs to be correct. 
%Unfortunately, while Alloy offers a concise formulation of complex properties, Alloy's support for expressive operators, such as transitive closure, can make writing non-trivial properties challenging.
In Alloy, there are two types of faults that can appear in a model: (1)~\emph{underconstrained} faults in which the model allows scenarios it should prevent, and (2)~\emph{overconstrained} faults in which the model prevents scenarios it should allow. 
%These faults can occur in isolation or combined. 
To detect an underconstrained fault, during enumeration, the user needs to observe a scenario that they did not expect to see. Unfortunately, even if the user notices something is wrong with a scenario, according to a recent user study, both novice and expert Alloy modelers struggled to refine a faulty formula given an incorrectly produced scenario~\cite{alloyformalise23study}. To detect an overconstrained fault, after enumeration, the user needs to realize that an expected scenario was never present. Since Alloy commands can produce hundreds of scenarios, this places a high burden on the user to realize their Alloy model is incorrect.

\usetikzlibrary{shapes.geometric, arrows}
\tikzstyle{arrow} = [line width=1.5pt,->,>=stealth]
\tikzstyle{darrow} = [line width=1.5pt,<->,>=stealth]
\tikzstyle{ListAtom} = [rectangle, minimum width=1.25cm, minimum height=.7cm, text centered, draw=black, fill=orange!75!yellow]
\tikzstyle{Trash} = [rectangle, minimum width=1.25cm, minimum height=.7cm, text centered, draw=black, fill=white!50!red]
\tikzstyle{Protected} = [rectangle, minimum width=1.25cm, minimum height=.7cm, text centered, draw=black, fill=yellow!75!orange]
\tikzstyle{State} = [circle, minimum width=0.5cm, minimum height=0.5cm, text centered, draw=black, fill=white]

\begin{figure*}
\begin{center}
\begin{tabular}[t]{c|c|c}
\footnotesize (a) & \multicolumn{2}{c}{\footnotesize (b)} \\
\begin{minipage}[t]{.6\columnwidth}
\scriptsize
\begin{Verbatim}[]
1. \Blue{var sig} File \{ \Blue{var} link : \Blue{lone} File \}
2. \Blue{var sig} Trash \Blue{in} File \{\}
3. \Blue{var sig} Protected \Blue{in} File \{\}
4. \Blue{pred} ProtectedNeverTrashed \{
5.    \Blue{no} Protected & Trash \Green{--incorrect}
6.    \Green{ --always no Protected & Trash --correct}
7. \}
8. \Blue{run } ProtectedNeverTrashed \Blue{for} \Red{3}
\end{Verbatim}
\end{minipage}

&

\begin{minipage}[t]{.45\columnwidth}
\begin{center}
\footnotesize
\begin{tikzpicture}[baseline,node distance=1.5cm]
\node (F0) [ListAtom] {\textbf{F0}};
\node (F1) [Trash, right of=F0] [align=center]{\textbf{F1}\\Trash};
\node (F2) [Protected, below of=F0] [align=center]{\textbf{F2}\\Protected};
\begin{scope}[brown]
\draw [darrow] (F0) -- node[anchor=west] {\textbf{link}} (F2);
\end{scope}[brown]
\end{tikzpicture}
\end{center}
\end{minipage}

&  
\begin{minipage}[t]{.5\columnwidth}
\begin{center}
\footnotesize
\begin{tikzpicture}[baseline,node distance=1.5cm]
\node (F0) [ListAtom] {\textbf{F0}};
\node  [xshift=3.5mm,rectangle,right of=F0,draw,text width=1.9cm,minimum height=.7cm,
        text centered,rounded corners,name = re] {};
       \filldraw[yellow!75!orange][] (re.south west)
        [] -- (re.south east)
        [] -- (re.north east)--cycle
        ;
        \filldraw[white!50!red][] (re.south west)
        [] -- (re.north west)
        [] -- (re.north east)--cycle
        ;
\node (F1) [xshift=3.5mm, rectangle,right of=F0, draw, text width=1.9cm,minimum height=.7cm,
        text centered,name = re] {\textbf{F1}\\Trash, Protected};     
\node (F2) [Protected, below of=F0] [align=center]{\textbf{F2}\\Protected};
\begin{scope}[brown]
\draw [darrow] (F0) -- node[anchor=west] {\textbf{link}} (F2);
\end{scope}[brown]
\end{tikzpicture}
\end{center}
\end{minipage}

\\
& \textbf{(State 0)} & \textbf{(State 1)}
\\
&
\multicolumn{2}{c}{
\begin{minipage}[t]{.45\columnwidth}
\begin{center}
\footnotesize
\begin{tikzpicture}[baseline,node distance=1cm]
\node (F0) [State] {\textbf{0}};
\node (F1) [State, right of=F0] {\textbf{1}};
\begin{scope}[black]
\draw [arrow] (F0) edge [line width=1pt] node {\textbf{}} (F1);
\draw [arrow] (F1) edge [loop right,line width=1pt,anchor=south] node {} (F1);
\end{scope}[black]
\end{tikzpicture}
\end{center}
\end{minipage}
}
\end{tabular}

\end{center}
\caption{Faulty Submission of an Alloy Model of a File System Trash Can}
\label{fig:trash}
\end{figure*}

Given these difficulties, there is a growing body of work looking at how to detect and debug faulty Alloy models. Initially, a unit testing framework AUnit, was created~\cite{AUnitConcept,AUnit}. AUnit enables users to outline a specific scenario they expect their model to allow or prevent and then check that this behavior actually occurs. %, which allows the user to directly check for under- or over-constrained faults without enumerating scenarios. 
AUnit has been used for mutation testing~\cite{MuAlloyTool,MuAlloyTemporal}, fault localization~\cite{AlloyFL}, and automated repair~\cite{ARepair,icebar}. Since then, fault localization and repair of Alloy models has become an active research area~\cite{atr,flack,beafix,tar}. 
As this field grows, we believe it is important to understand what mistakes new developers actually make when writing models so that these techniques can (1) be effective for the majority of mistakes made, (2) be evaluated with respect to real faults and not simulated faults and (3) be properly leveraged to ease the burden of learning to write software models. To achieve these goals, this paper presents an empirical study that explores the common development practices of novice Alloy modelers using 97,755 models submitted through Alloy4Fun, an online educational website for Alloy. %There are 17 models and 97,755 submissions. 

In this paper, we make the following contributions: 

\newenvironment{Contributions}{}{} %{\begin{itemize}}{\end{itemize}}
\newcommand{\Contribution}[1]{\textbf{#1:}} %{\item \textbf{#1.}}

\begin{Contributions}

\Contribution{Empirical Study} We present a systematic study of both correct and incorrect models written by novice Alloy users with revision histories that capture back-to-back incremental changes. 

\Contribution{Practical Impacts} We distill our observations into practical guidelines for future work in debugging models and building educational material for formal methods. 

\Contribution{Benchmarks for Alloy} As part of exploration, we create two benchmarks: (1) a collection of real-world faulty models broken down by type of fault and (2) a collection of models that tracks changes, which can be used to evaluate and improve incremental analysis techniques for Alloy~\cite{Reach,iAlloy,titanium,Platinum}.

%\Contribution{Open Source} We release our benchmarks and analysis at: \url{https://osf.io/93c6r/?view_only=e8b878b16ae4456a9a1688a03d5b659a}
\Contribution{Open Source} We release our benchmarks and data analysis at: \url{https://github.com/NoviceAlloyModels/NoviceAlloyStudy}.

\end{Contributions}

\section{Background}\label{sec:example}

In this section, we describe the key concepts of Alloy and Alloy4Fun.

\subsection{Alloy}

Figure~\ref{fig:trash} (a) displays a faulty temporal model of a file system trash can from the Alloy4Fun dataset~\cite{alloy4funbenchmark}. Signature paragraphs introduce named sets and can define relations, which outline relationships between elements of sets. Line 1 introduces a named set \CodeIn{File} and establishes that each \CodeIn{File} atom connects to zero or one (\CodeIn{lone}) \CodeIn{File} atoms through the \CodeIn{link} relation. Lines 2 and 3 introduce the named sets \CodeIn{Trash} and \CodeIn{Protected} as subsets (\CodeIn{in}) of \CodeIn{File}.  Signatures and relations can be declared mutable (\CodeIn{var}), which means that the elements of these sets can vary across different states in the same scenario. In our example, all signatures and relations are mutable.

Predicate paragraphs introduce named formulas that can be invoked elsewhere. The predicate \CodeIn{ProtectedNotTrashed} uses empty set (`\CodeIn{no}') and set intersection (`\CodeIn{\&}') to incorrectly attempt to establish that a protected file is never sent to the trash. However, since the signatures are mutable, the incorrect version is true as long as no protected files are in the trash for the \emph{first} state but does not require the constraint to be true in \emph{every} state. To correct this, the linear temporal operator \CodeIn{always} can be appended to the start. Commands indicate which formulas to invoke and what scope to explore. The scope places an upper bound on the size of all signature sets and the number of state transitions. The command on line 8 instructs the Analyzer to search for an assignment to all sets in the model using up to 3 \CodeIn{File} atoms and up to 10 state transitions by default. Figure~\ref{fig:trash} (a) displays the first scenario found by the Analyzer, which informs the user that their model is incorrect: in state 1, \CodeIn{F1} is both in the trash and protected. %At this point, the user can return to their model, edit their constraints, and re-run the command until all scenarios enumerated match their expectation.

\subsection{Alloy4Fun Exercises} 
Alloy4Fun is an online educational platform for learning Alloy where users can write and compile Alloy models through an online interface. %, without having to install the Analyzer. 
Alloy4Fun contains several starter models in which the signature paragraphs are given. Each starter model contains a varying number of empty predicates with a corresponding English description of the property that the user can attempt to encode into formal logic. To check their answer for a given predicate, Alloy4Fun will compare if their written submission is equivalent to a hidden oracle. If their submission is not equivalent to the oracle, the user will be presented with a counterexample: a scenario that depicts a situation where the user's formula behaves differently than the oracle. The user can then continue to iterate on their formula until it is correct. To illustrate, Figure~\ref{fig:trash_a4fun} displays the Trash exercise that generated the model in Figure~\ref{fig:trash}. Lines 1-2 are visible to the user while lines 3-7 are hidden. The check command (lines 5-7) compares if a submission is equivalent to the oracle, which checks for semantic equivalence and not syntactic equivalence. 

\usetikzlibrary{shapes.geometric, arrows}
\tikzstyle{arrow} = [line width=1.5pt,->,>=stealth]
\tikzstyle{darrow} = [line width=1.5pt,<->,>=stealth]
\tikzstyle{ListAtom} = [rectangle, minimum width=1.5cm, minimum height=.75cm, text centered, draw=black, fill=orange!75!yellow]
\tikzstyle{Trash} = [rectangle, minimum width=1.5cm, minimum height=.75cm, text centered, draw=black, fill=white!50!red]
\tikzstyle{Protected} = [rectangle, minimum width=1.5cm, minimum height=.75cm, text centered, draw=black, fill=yellow!75!orange]
\tikzstyle{State} = [circle, minimum width=0.5cm, minimum height=0.5cm, text centered, draw=black, fill=white]

\begin{figure}
\begin{minipage}[t]{.8\columnwidth}
\scriptsize
\begin{Verbatim}[]
1. \Green{// a protected file is at no time sent to the trash}
2. \Blue{pred} prop9 \{\}
3. \Green{//SECRET}
4. \Blue{pre}d prop9o \{ \Blue{always no} Protected & Trash \}
5. \Blue{check} prop9Ok \{ 
6.    (prop9 \Blue{and not} prop9o) \Blue{implies} (prop9o \Blue{iff} prop9) 
7. \}
\end{Verbatim}
\end{minipage}
\caption{Snipet from the Trash Alloy4Fun Exercise}
\label{fig:trash_a4fun}
\end{figure}

% 1. \Blue{var sig} File \{ \Blue{var} link : \Blue{lone} File \}
% 2. \Blue{var sig} Trash \Blue{in} File \{\}
% 3. \Blue{var sig} Protected \Blue{in} File \{\}

%This dataset contains models submitted by students in the Alloy4Fun platform to solve the challenge models from various editions of formal methods courses in the University of Minho (UM) and the University of Porto (UP) between the fall of 2019 and the spring of 2023, totaling about 100.000 entries. Participants include those enrolled in the optional MSc course "Specification and Modelling" (EM) and the mandatory MSc course "Formal Methods in Software Engineering" (MFES) in UM, and the optional MSc course "Formal Methods for Critical Systems" (MFS) in UP. Note that since the challenges' permalinks are publicly available, the dataset may contain submissions from other participants outside the classroom context.
\section{Experiment SetUp}

In this paper, we will use the following terminology. A \textit{submission} consists of the base model, the user's current attempt for a specific exercise, and any helper predicates invoked. An \textit{edit path} is the series of submissions a user makes starting with the starter model and ending with their last submission. Edit paths can contain submissions across multiple different exercises but are all within the same model. An \textit{attempt} is the series of submissions within an edit path that the user makes for a specific exercise. %An attempt is a sub-portion of an edit path.

\subsection{Experimental Data}

In our study, we use the publicly released Alloy4Fun dataset, which contains real-world models obtained from student submissions in the academic period from Fall 2019 to Spring 2023~\cite{alloy4funbenchmark}. In total, there are 97,755 submissions that span 17 different Alloy models and 183 predicates to be filled in (exercises). Of these, we filtered out 2,024 submissions that change the model's structure or are only changes to the theme settings for visualization. In addition, we filtered 2,448 submissions that are empty. Empty submissions are underconstrained faults. However, since empty submissions do not fundamentally illustrate any effort by the novice user, we removed them to avoid skewing the conclusions. As a result, we end up with 93,283 submissions, all of which have tangible user-created content. %Since users can only check one predicate at a type, we consider a submission to the starter model plus the checked predicate and ignore any other predicates in the model not directly invoked by the checked predicate.

Table~\ref{tab:model_stats} gives an overview of the complexity of the models used in our study in terms of the universe of discourse each model creates. Column \textbf{\#Sig} is the total number of signatures in the model, \textbf{\#Abs} is the number of abstract signatures, \textbf{\#Ext} is the number of signatures that extend another signature, \textbf{\#Rel} is the number of relations, \textbf{Arity} is the average arity of all relations in the model (2 indicates a binary relation), and \textbf{\#Exe} is the number of exercises. Models with an underscore in their name represent models that have multiple versions in the dataset. Between versions, the number of exercises, instructional text, and/or the type of logic can change. 

\begin{table}
  \centering
\caption{Overview of Complexity of Models in Study}
  \label{tab:model_stats}
  \resizebox{!}{.38\columnwidth}{
  \begin{tabular}{lllllll} \hline
\rowcolor[HTML]{C0C0C0} 
 & \textbf{\#Sig} & \textbf{\#Abs} & \textbf{\#Ext} & \textbf{\#Rel} & \textbf{Arity} & \textbf{\#Exe} \\ \hline
\textbf{classroom\_fol} & 5 & 0 & 2 & 3 & 2.33 & 15 \\ 
\rowcolor[HTML]{EFEFEF} 
\textbf{classroom\_rl} & 5 & 0 & 2 & 3 & 2.33 & 15 \\
\textbf{courses\_v1} & 6 & 0 & 2 & 5 & 2.2 & 15 \\
\rowcolor[HTML]{EFEFEF} 
\textbf{courses\_v2} & 6 & 0 & 2 & 5 & 2.2 & 15 \\
\textbf{cv\_v1} & 5 & 1 & 2 & 4 & 2 & 4 \\
\rowcolor[HTML]{EFEFEF} 
\textbf{cv\_v2} & 5 & 1 & 2 & 4 & 2 & 4 \\
\textbf{graphs} & 1 & 0 & 0 & 1 & 2 & 8 \\
\rowcolor[HTML]{EFEFEF} 
\textbf{lts} & 3 & 0 & 1 & 1 & 3 & 6 \\
\textbf{productionLine\_v1} & 5 & 0 & 2 & 3 & 2 & 4 \\
\rowcolor[HTML]{EFEFEF} 
\textbf{productionLine\_v2} & 10 & 1 & 7 & 4 & 2 & 10 \\
\textbf{productionLine\_v3} & 10 & 1 & 7 & 4 & 2 & 10 \\
\rowcolor[HTML]{EFEFEF} 
\textbf{socialMedia} & 5 & 0 & 2 & 5 & 2 & 8 \\
\textbf{trainstation\_fol} & 7 & 0 & 5 & 2 & 2 & 10 \\
\rowcolor[HTML]{EFEFEF} 
\textbf{trainstation\_ltl} & 6 & 0 & 4 & 3 & 2 & 17 \\
\textbf{trash\_rl} & 3 & 0 & 2 & 1 & 2 & 10 \\
\rowcolor[HTML]{EFEFEF} 
\textbf{trash\_ltl} & 3 & 0 & 2 & 1 & 2 & 20 \\
\textbf{trash\_fol} & 3 & 0 & 2 & 1 & 2 & 10 \\ \hline
\rowcolor[HTML]{DAE8FC}  
\textbf{AVG} & \textbf{5.18} & \textbf{0.24} & \textbf{2.71} & \textbf{2.94} & \textbf{2.12} & \textbf{10.65} \\ \hline
\end{tabular}
  }
\end{table}

An exercise falls into one of the following categories:  relational logic (RL), predicate logic (PL), first-order logic (FOL), and linear temporal logic (LTL). %To illustrate the bounds between the different types of logic, 
The following is the subset of Alloy's grammar that captures all supported relational logic operators:

\begin{footnotesize}
\begin{Verbatim}[frame=lines,rulecolor=\color{lightgray}]
unOp::= ~ | * | ^ | ! | not | no | mult | set | 
binOp::= & | + | - | ++ | <: | :> | .
arrowOp::= [mult | set] -> [mult | set]
compareOp::= in | = | < | > | =< | >=
expr::= { decl,+ blockOrBar } (comprehension)
mult::= lone | some | one
\end{Verbatim}
\end{footnotesize}

\noindent The following is the subset of predicate logic operators:
\begin{footnotesize}
\begin{Verbatim}[frame=lines,rulecolor=\color{lightgray}]
binOp::= || | or | && | and | <=> | iff | => | implies
\end{Verbatim}
\end{footnotesize}

\noindent The following is the subset of first-order logic operators:
\begin{footnotesize}
\begin{Verbatim}[frame=lines,rulecolor=\color{lightgray}]
quant::= all | no | sum | mult
expr::= quant decl,+ blockOrBar
\end{Verbatim}
\end{footnotesize}

\noindent The following is the subset of linear temporal logic operators:
\begin{footnotesize}
\begin{Verbatim}[frame=lines,rulecolor=\color{lightgray}]
unOp::= always | eventually | after | before | historically | once | '
binOp::=  since | triggered | until | releases | ;
\end{Verbatim}
\end{footnotesize}

Table~\ref{tab:exe_stats} conveys the complexity of the exercise based on the type of logic. Exercises were classified based on the highest operator present in the oracle solution according to the partial order of RL$\rightarrow$PL$\rightarrow$FOL$\rightarrow$LTL. 
%Of note, RL and PL exercises can be solved using FOL and sometimes, FOL exercises can be solved using RL and PL. 
After \textbf{Type}, Column \textbf{\#Exe} is the number of exercises of that type, and \textbf{Rate} is the percentage of total exercises that are the associated type. The last 5 columns demonstrate the size of the oracle solution broken down by type of logic. Column \textbf{\#AST} is the average number of AST nodes in the oracle, and columns \textbf{\#O$_{RL}$}, \textbf{\#O$_{PL}$}, \textbf{\#O$_{FOL}$}, and \textbf{\#O$_{LTL}$} are the average number of operators that appear of each type of logic that appears in the oracle solutions.

\subsection{Classification of Submissions}
Within our dataset, a submission falls into one of the following categories: correct, overconstrained (over), underconstrained (under), both over- and underconstrained (both), type error (type), and syntax error (syntax).  Correct submissions are semantically equivalent to the oracle, overconstrained submissions prevent valid scenarios from being generated, underconstrained submissions allow invalid scenarios to be generated, both submissions allow a combination of over- and underconstrained behaviors, type errors are formulas that are syntactically valid but fail to typecheck, and syntax errors are formulas that fail to adhere to the grammar rules.

\begin{table}
  \centering
\caption{Number and Size of Exercises by Type}
  \label{tab:exe_stats}
  \resizebox{!}{.12\columnwidth}{
  \begin{tabular}{lrr|rrrrr} \hline
\rowcolor[HTML]{C0C0C0} 
\multicolumn{1}{c}{\cellcolor[HTML]{C0C0C0}\textbf{Type}} & \multicolumn{1}{c}{\cellcolor[HTML]{C0C0C0}\textbf{\# Exe}} & \multicolumn{1}{c|}{\cellcolor[HTML]{C0C0C0}\textbf{Rate}} & \multicolumn{1}{c}{\cellcolor[HTML]{C0C0C0}\textbf{\#AST}} & \multicolumn{1}{c}{\cellcolor[HTML]{C0C0C0}\textbf{\#O$_{RL}$}} & \multicolumn{1}{c}{\cellcolor[HTML]{C0C0C0}\textbf{\#O$_{PL}$}} & \multicolumn{1}{c}{\cellcolor[HTML]{C0C0C0}\textbf{\#O$_{FOL}$}} & \multicolumn{1}{c}{\cellcolor[HTML]{C0C0C0}\textbf{\#O$_{LTL}$}} \\ \hline
\textbf{RL} & 51 & 27.87 & 4.18 & 1.84 & 0.00 & 0.00 & 0 \\
\rowcolor[HTML]{EFEFEF} 
\textbf{PL} & 1 & 0.55 & 5.00 & 2.00 & 1.00 & 0.00 & 0 \\
\textbf{FOL} & 95 & 51.91 & 17.47 & 5.88 & 0.31 & 1.07 & 0 \\
\rowcolor[HTML]{EFEFEF} 
\textbf{LTL} & 36 & 19.67 & 16.31 & 4.72 & 0.56 & 0.81 & 2 \\ \hline
\rowcolor[HTML]{DAE8FC} 
\textbf{SUM/AVG} & \textbf{183} & \textbf{100.00} & \textbf{10.74} & \textbf{3.61} & \textbf{0.47} & \textbf{0.47} & \textbf{0.50} \\ \hline
\end{tabular}}
\end{table}

\begin{figure*}
    \centering
    \usetikzlibrary{angles,calc,positioning}
\begin{tabular}{ccccc}

\begin{tikzpicture}
\pie[color={blue!60!cyan!60, yellow!60, orange!60, red!60, yellow!70!cyan!80, purple!80!cyan!60},radius=1.1,hide number]{25.40/\textbf{}, 8.68/\textbf{}, 23.28/\textbf{}, 13.36/\textbf{}, 14.65/\textbf{}, 14.64/\textbf{}}

\pie[color={blue!60!cyan!60,  yellow!60, orange!60, red!60, yellow!70!cyan!80, purple!80!cyan!60}, pos={3,0},radius=1.1,hide number]{37.05/\textbf{}, 9.15/\textbf{}, 17.01/\textbf{}, 7.38/\textbf{}, 16.81/\textbf{}, 12.6/\textbf{}}

\pie[color={blue!60!cyan!60,  yellow!60, orange!60, red!60, yellow!70!cyan!80, purple!80!cyan!60}, pos={6,0},radius=1.1,hide number]{25.18/\textbf{}, 11.43/\textbf{},  18.98/\textbf{}, 6.78/\textbf{}, 22.98/\textbf{}, 14.65/\textbf{}}

\pie[color={blue!60!cyan!60,  yellow!60, orange!60, red!60, yellow!70!cyan!80, purple!80!cyan!60}, pos={9,0},radius=1.1,hide number]{23.62/\textbf{}, 8.27/\textbf{}, 23.88/\textbf{}, 14.22/\textbf{}, 14.79/\textbf{}, 15.21/\textbf{}}

\pie[color={blue!60!cyan!60,  yellow!60, orange!60, red!60, yellow!70!cyan!80, purple!80!cyan!60}, pos={12,0},radius=1.1,hide number,text = legend]{24.13/\textbf{Corr},  11.93/\textbf{Over}, 29.21/\textbf{Both}, 16.19/\textbf{Under}, 6.8/\textbf{Type}, 11.74/\textbf{Syntax}}

\node[yshift=-1.5cm,text=black] (start) { \textbf{(ALL)} };
\node[yshift=-1.5cm,xshift=3cm,text=black] (start) { \textbf{(RL)} };
\node[yshift=-1.5cm,xshift=6cm,text=black] (start) { \textbf{(PL)} };
\node[yshift=-1.5cm,xshift=9cm,text=black] (start) { \textbf{(FOL)} };
\node[yshift=-1.5cm,xshift=12.1cm,text=black] (start) { \textbf{(LTL)} };

\end{tikzpicture}

\end{tabular}
    \caption{Breakdown of Submission Results}
    \label{fig:cx_sub}
\end{figure*}

\subsection{Recreating Submission Revision Histories}

%For the purpose of gaining better insights into what are the steps that the novice users are performing to get to the correct solution, we have created submission history trees using the student submissions. 
When a student checks an exercise, Alloy4Fun logs the student submission with a unique id key and a ``derivationOf'' parameter that contains the unique id of the parent entry. If the submission is the first attempt, the parent model is the officially published Alloy4Fun example model. Otherwise, the parent is the previous submission made by the user. We use this information to re-build edit paths that capture the series of edits a user makes. As an example, the following is the edit path for our example in Figure~\ref{fig:trash}:

\begin{footnotesize}
\begin{Verbatim}[frame=lines,rulecolor=\color{lightgray}]
inv9 U: \Blue{no} Protected & Trash
inv9 C: \Blue{always no} Protected & Trash
\end{Verbatim}
\end{footnotesize}

For edit paths, we do include empty submissions. While empty submissions do not hold much value as individual submissions, in the middle of an edit path, an empty submission can convey when users reset their attempt on an exercise.

\section{Empirical Evaluation}\label{sec:eval}

%Our research questions are organized by explorations of different classification of submissions. 
%We first explore general behavior across all submissions (RQ1-RQ3) before exploring how users write invalid submissions (RQ4), correct submissions (RQ5-RQ6), and incorrect submissions (RQ7-RQ8). Finally, we consider how users incrementally problem solve (RQ9). 

We first explore general behavior across all submissions (RQ1-RQ2) before exploring how users write invalid submissions (RQ3), correct submissions (RQ4-RQ5), and incorrect submissions (RQ6-RQ7). Finally, we consider how users incrementally problem solve (RQ8).

\subsection{RQ1: What classification of submissions do novice users make and at what rate?}

To get a general overview of the submissions made, Figure~\ref{fig:cx_sub} shows pie charts depicting the breakdown in the classification of submissions. %In subsequent RQs, we dive deeper into the submissions within these classifications.
The first pie chart contains information for all submissions, while the remaining shows the breakdown per type of logic. Submissions were sorted into different types of logic based on the logic in the user's submission and not the backend oracle solution. 

As the pie charts indicate, novice users are the most effective at writing relational logic properties, which novices express correctly 37.05\% of the time. In contrast, novice users struggle more with FOL and LTL properties, in which less than a fourth of the submissions are correct: 23.62\% and 24.13\% respectively. This decrease in accuracy for FOL and LTL exercises is expected, as the introduction of quantified formulas and temporal constraints is non-trivial. 

While correct submissions are the largest individual chunk at 25.4\%, faulty submissions (both, over and under) combine to account for 45.3\% of all submissions. Faults that are both over- and underconstrained account for the largest portion of mistakes at 23.28\% of all submissions. However, the distribution of strictly overconstrained or strictly underconstrained faults varies by logic. The rate of underconstrained formulas notably increases for FOL and LTL submissions, going from around 7\% for RL and PL to about 15\% for FOL and LTL submissions. This indicates that the addition of quantification and temporal logic results in novice users writing properties that are often too permissive. In latter questions, we dive deeper into why different types of faults occur. 

Invalid submissions combine to account for 29.28\% of all submissions and are almost evenly split between syntax errors (14.64\%) and type errors (14.65\%). While the rate of syntax errors is relatively consistent across types of logic, the rate of type errors varies greatly. Over 15\% of RL, PL, and FOL submissions have type errors, while only 7\% of LTL submissions have type errors. One possible explanation for this is that within the LTL exercises, only some of the signatures and relations are mutable. As a result, there is a more narrow pool of types for which temporal properties apply, resulting in novices making fewer mistakes.

\begin{impact}
While FOL and LTL properties are more difficult, the reduction in type errors for LTL exercises gives insight into ways to ease teaching these logics. For instance, educators can highlight how FOL quantified domains also narrow the pool of types.
%FOL and LTL properties are more difficult for novice users. Given that there are fewer type errors for LTL exercises, educators can highlight how the type introduced in a quantified domain reduces the applicable types as well.
%Novice users have more difficulty with FOL and LTL properties. However, LTL exercises tend to reduce the number of type errors, as only subsets of the model are mutable.
%Novice users are also more likely to write formulas that are too permissive; therefore, debugging efforts should focus on additive constraints.
\end{impact}

\begin{table}
\centering
\caption{Rate of Syntactically Unique Submissions}
\resizebox{!}{.35\columnwidth}{
\begin{tabular}{lrrr||lrrr} \hline
\rowcolor[HTML]{C0C0C0} 
\multicolumn{4}{c||}{\cellcolor[HTML]{C0C0C0}\textbf{RL}} & \multicolumn{4}{c}{\cellcolor[HTML]{C0C0C0}\textbf{PL}} \\
\rowcolor[HTML]{C0C0C0} 
\multicolumn{1}{c}{\cellcolor[HTML]{C0C0C0}\textbf{Clx}} & \multicolumn{1}{c}{\cellcolor[HTML]{C0C0C0}\textbf{\#Sub}} & \multicolumn{1}{c}{\cellcolor[HTML]{C0C0C0}\textbf{\#Uni}} & \multicolumn{1}{c||}{\cellcolor[HTML]{C0C0C0}\textbf{\%Uni}} & \multicolumn{1}{c}{\cellcolor[HTML]{C0C0C0}\textbf{Clx}} & \multicolumn{1}{c}{\cellcolor[HTML]{C0C0C0}\textbf{\#Sub}} & \multicolumn{1}{c}{\cellcolor[HTML]{C0C0C0}\textbf{\#Uni}} & \multicolumn{1}{c}{\cellcolor[HTML]{C0C0C0}\textbf{\%Uni}} \\ \hline
\textbf{Correct} & 4418 & 397 & 15.70 & \textbf{Correct} & 390 & 60 & 13.93 \\
\rowcolor[HTML]{EFEFEF} 
\textbf{Both} & 2029 & 1139 & 62.32 & \textbf{Both} & 294 & 208 & 72.49 \\
\textbf{Over} & 1091 & 558 & 51.52 & \textbf{Over} & 177 & 100 & 44.00 \\
\rowcolor[HTML]{EFEFEF} 
\textbf{Under} & 880 & 434 & 65.89 & \textbf{Under} & 105 & 77 & 80.85 \\
\textbf{Syntax} & 1503 & 1150 & 84.88 & \textbf{Syntax} & 227 & 199 & 78.63 \\
\rowcolor[HTML]{EFEFEF} 
\textbf{Type} & 2004 & 1460 & 77.07 & \textbf{Type} & 356 & 279 & 83.64 \\ \hline
\rowcolor[HTML]{DAE8FC} 
\textbf{SUM/AVG} & \textbf{11925} & \textbf{5138} & \textbf{43.09} & \textbf{SUM/AVG} & \textbf{1549} & \textbf{923} & \textbf{59.59} \\ \hline\hline
\rowcolor[HTML]{C0C0C0} 
\multicolumn{4}{c||}{\cellcolor[HTML]{C0C0C0}\textbf{FOL}} & \multicolumn{4}{c}{\cellcolor[HTML]{C0C0C0}\textbf{LTL}} \\
\rowcolor[HTML]{C0C0C0} 
\multicolumn{1}{c}{\cellcolor[HTML]{C0C0C0}\textbf{Clx}} & \multicolumn{1}{c}{\cellcolor[HTML]{C0C0C0}\textbf{\#Sub}} & \multicolumn{1}{c}{\cellcolor[HTML]{C0C0C0}\textbf{\#Uni}} & \multicolumn{1}{c||}{\cellcolor[HTML]{C0C0C0}\textbf{\%Uni}} & \multicolumn{1}{c}{\cellcolor[HTML]{C0C0C0}\textbf{Clx}} & \multicolumn{1}{c}{\cellcolor[HTML]{C0C0C0}\textbf{\#Sub}} & \multicolumn{1}{c}{\cellcolor[HTML]{C0C0C0}\textbf{\#Uni}} & \multicolumn{1}{c}{\cellcolor[HTML]{C0C0C0}\textbf{\%Uni}} \\ \hline
\textbf{Correct} & 17373 & 3800 & 21.85 & \textbf{Correct} & 1513 & 205 & 12.58 \\
\rowcolor[HTML]{EFEFEF} 
\textbf{Both} & 17561 & 11592 & 65.47 & \textbf{Both} & 1831 & 1191 & 64.70 \\
\textbf{Over} & 6081 & 3672 & 61.61 & \textbf{Over} & 748 & 420 & 53.64 \\
\rowcolor[HTML]{EFEFEF} 
\textbf{Under} & 10459 & 5422 & 50.06 & \textbf{Under} & 1015 & 522 & 50.68 \\
\textbf{Syntax} & 11188 & 9361 & 82.50 & \textbf{Syntax} & 736 & 643 & 87.03 \\
\rowcolor[HTML]{EFEFEF} 
\textbf{Type} & 10878 & 8596 & 78.33 & \textbf{Type} & 426 & 330 & 76.64 \\ \hline
\rowcolor[HTML]{DAE8FC} 
\textbf{SUM/AVG} & \textbf{73540} & \textbf{42443} & \textbf{57.71} & \textbf{SUM/AVG} & \textbf{6269} & \textbf{3311} & \textbf{52.82} \\ \hline
\end{tabular}
}
\label{tab:unique_type_detailed}
\end{table}

\subsection{RQ2: How often do novice users make the same submission?}

If users often repeat the same submission, then this represents underlying common approaches that novices take to modeling, and understanding what these common approaches are can help guide improvements in tool support and educational material. Therefore, we investigate how often new users make duplicate submissions. We consider two forms of equivalence: syntactic and semantic. 
%Syntactical duplicates covey the rate at which novice users literally write the same exact formula, while semantic duplicates allow us to explore the rate at which users express the same underlying concept. 
%Frequently occurring semantic mistakes can represent a fundamental misunderstanding that could be happening, which reflects an educational gap and/or ambiguous wording on the instructions for the exercise. 
To determine if two submissions are syntactically equivalent, we used a PrettyStringVisitor to reprint formulas in a consistent format that eliminates trivial differences in formatting that arise from different uses of white spaces. To determine if two submissions are semantically equivalent, we used the Analyzer to check for logical equivalence between submissions.  

Table~\ref{tab:unique_type_detailed} and Table~\ref{tab:semantic} show the rate at which submissions are syntactically and semantically unique broken down by classification (\textbf{Clx}) and additionally type of logic for syntactic. For both tables, Column \textbf{\#Sub} is the total number of submissions, column \textbf{\#Uni} is the number of submissions that are unique syntactically or semantically, \textbf{\%Uni} is the percentage of total submissions that are unique. Syntax and Type submissions do not have semantically equivalent values since these submissions do not compile. 

In general, novices do have common approaches to modeling -- only 55.55\% of submissions are syntactically unique and 16.55\% are semantically unique. Two main trends impact this. First, there is high redundancy in correct submissions, which is expected as there are a finite number of ways to express a property correctly. Moreover, semantically, all correct submissions for an exercise are equivalent. 
%For correct submissions, the number of unique semantic submissions is equivalent to the number of exercises that have at least one correct submission, which is expected as all correct solutions should be semantically equivalent. 
Second, there is notably less redundancy in type and syntax errors, meaning that novice users make a broader range of mistakes when incorrectly structuring a formula compared to writing a syntactically valid formula that is wrong.

For faulty submissions, RL and PL submissions tend to make the same overly restrictive formula (over) mistakes syntactically, while FOL and LTL submissions tend to repeat the same overly permissive mistakes (under). As RQ6 will further highlight, one reason for this split phenomenon is that when novice users write incorrect relational formulas, they tend to write a formula that checks for a concept too broadly across an entire set of atoms when it should be checked per atom of a set, overconstraining the model. In contrast, when writing a quantified formula, the user will apply a formula to too narrow of a domain, underconstraining the model. % while for linear temporal logic formulas, users over leave out or select the wrong operator also underconstraining the model. 
Semantically, users are almost always making the same underconstrained fault, as only 9.83\% of the formulas are unique logically. However, with at most 36.17\% of the formulas being unique across all classifications of mistakes, users often repeat the same high-level logical mistakes. 

\begin{table}
\centering
\caption{Rate of Semantically Unique Submissions}
\label{tab:semantic}
\begin{footnotesize}
\begin{tabular}{lrrr} \hline
\rowcolor[HTML]{C0C0C0} 
\multicolumn{1}{c}{\cellcolor[HTML]{C0C0C0}\textbf{Clx}} & \multicolumn{1}{c}{\cellcolor[HTML]{C0C0C0}\textbf{\#Sub}} & \multicolumn{1}{c}{\cellcolor[HTML]{C0C0C0}\textbf{\#Uni}} & \multicolumn{1}{c}{\cellcolor[HTML]{C0C0C0}\textbf{\%Uni}} \\ \hline
\textbf{Correct} & 23694 & 176 & 0.74 \\
\rowcolor[HTML]{EFEFEF} 
\textbf{Both} & 21715 & 7854 & 36.17 \\
\textbf{Over} & 8097 & 1659 & 20.49 \\
\rowcolor[HTML]{EFEFEF} 
\textbf{Under} & 12459 & 1225 & 9.83 \\ \hline
\rowcolor[HTML]{DAE8FC} 
\textbf{SUM/AVG} & \textbf{65965} & \textbf{10914} & \textbf{16.55} \\ \hline
\end{tabular}
\end{footnotesize}
\end{table}

\begin{impact}
Novice users have common approaches they take to express constraints. Given the high redundancy in underconstrained faults, a good starting point is for researchers and educators to focus on how to prevent too permissive of formulas. 
\end{impact}

%\begin{table}
%\centering
%\caption{Rate of Duplicate and Semantically Novel Submissions by Submission Result}
%\label{tab:unique_error}
%\begin{footnotesize}
%\input{tables/unique_to_dups_by_error}
%\end{footnotesize}
%\end{table}

%\begin{table}
%\centering
%\caption{Rate of Duplicate and Semantically Novel Submissions by Type of Logic}
%\label{tab:unique_type}
%\begin{footnotesize}
%\input{tables/unique_to_dups_type}
%\end{footnotesize}
%\end{table}

\subsection{RQ3: How effective is the Analyzer's compilation error reports?}

Nearly a third of the time, a novice user will create a submission that fails to compile. Therefore, we investigate the effectiveness of Alloy's current compiler-based error reporting, which uses its built-in parser plus an internal type system to report syntax and type errors to the user. This effort is summarized in Table~\ref{tab:invalid_atmps}.
%presents the details for all attempts to solve an exercise that has at least one invalid submission. 
For each model, column 2 reports the percentage of attempts that have at least one invalid submission, and column 3 presents the percentage of attempts that never turned into valid formulas. The remaining three columns aim to quantify how a user goes from an invalid submission to a valid submission. We first collected all invalid subpaths, which are the portions of attempts that consist of only type or syntax error edits. Then, column 4 reports the average length, column 5 reports the percentage of subpaths that have a length of 5 or more, and column 6 reports the max length of any subpath. Since an attempt often represents more than one submission and the user can switch between types of logic as they make edits, we do not report information based on the type of logic.

\begin{table}
\centering
\caption{Details for Invalid Attempts}
\resizebox{!}{.35\columnwidth}{
\begin{tabular}{l|rr|rrr} 
\hline
\rowcolor[HTML]{C0C0C0} 
\multicolumn{1}{c|}{\cellcolor[HTML]{C0C0C0}} & \multicolumn{1}{c}{\cellcolor[HTML]{C0C0C0}} & \multicolumn{1}{c|}{\cellcolor[HTML]{C0C0C0}} & \multicolumn{3}{c}{\cellcolor[HTML]{C0C0C0}\textbf{Invalid Subpath Length}} \\ \cline{4-6} 
\rowcolor[HTML]{C0C0C0} 
\multicolumn{1}{c|}{\multirow{-2}{*}{\cellcolor[HTML]{C0C0C0}\textbf{Model}}} & \multicolumn{1}{c}{\multirow{-2}{*}{\cellcolor[HTML]{C0C0C0}\textbf{\begin{tabular}[c]{@{}c@{}}\% Atmps \\ w/ Invalid\end{tabular}}}} & \multicolumn{1}{c|}{\multirow{-2}{*}{\cellcolor[HTML]{C0C0C0}\textbf{\begin{tabular}[c]{@{}c@{}}\% Atmps\\ Nvr Valid\end{tabular}}}} & \multicolumn{1}{c|}{\cellcolor[HTML]{C0C0C0}\textbf{Avg}} & \multicolumn{1}{c|}{\cellcolor[HTML]{C0C0C0}\textbf{Pct $\geq$ 5}} & \multicolumn{1}{c}{\cellcolor[HTML]{C0C0C0}\textbf{Max}} \\ \hline
\textbf{classroom\_fol} & 30.44 & 2.75 & 1.93 & 7.50 & 11 \\
\rowcolor[HTML]{EFEFEF} 
\textbf{classroom\_rl} & 38.68 & 4.44 & 2.04 & 8.63 & 15 \\
\textbf{courses\_v1} & 48.89 & 1.79 & 2.10 & 8.89 & 22 \\
\rowcolor[HTML]{EFEFEF} 
\textbf{courses\_v2} & 50.28 & 2.71 & 1.81 & 4.57 & 20 \\
\textbf{cv\_v1} & 41.90 & 5.93 & 1.81 & 2.91 & 18 \\
\rowcolor[HTML]{EFEFEF} 
\textbf{cv\_v2} & 43.48 & 0.00 & 1.87 & 5.66 & 9 \\
\textbf{graphs} & 26.18 & 2.09 & 1.76 & 4.29 & 9 \\
\rowcolor[HTML]{EFEFEF} 
\textbf{lts} & 40.95 & 8.62 & 2.34 & 10.11 & 17 \\
\textbf{productionLine\_v1} & 32.91 & 1.28 & 2.10 & 8.77 & 15 \\
\rowcolor[HTML]{EFEFEF} 
\textbf{productionLive\_v2} & 41.43 & 1.00 & 1.84 & 5.15 & 37 \\
\textbf{productionLine\_v3} & 36.38 & 1.54 & 1.84 & 4.65 & 11 \\
\rowcolor[HTML]{EFEFEF} 
\textbf{socialMedia} & 48.27 & 4.19 & 2.06 & 8.56 & 23 \\
\textbf{trainStation\_fol} & 50.25 & 0.98 & 1.91 & 6.20 & 11 \\
\rowcolor[HTML]{EFEFEF} 
\textbf{trianStation\_ltl} & 37.73 & 4.09 & 1.98 & 6.85 & 12 \\
\textbf{trash\_fol} & 26.40 & 4.40 & 1.96 & 8.67 & 11 \\
\rowcolor[HTML]{EFEFEF} 
\textbf{trash\_ltl} & 32.35 & 0.67 & 1.91 & 7.71 & 9 \\
\textbf{trash\_rl} & 28.81 & 3.17 & 2.10 & 8.67 & 15 \\ \hline
\rowcolor[HTML]{DAE8FC}  
\textbf{SUM/AVG} & \textbf{42.06} & \textbf{2.88} & \textbf{1.99} & \textbf{7.48} & \textbf{37} \\ \hline
\end{tabular}
}
\label{tab:invalid_atmps}
\end{table}

Overall, the error reports help novices most of the time. On average, users correct an error within 2 attempts. However, there are gaps in the effectiveness of the current reports: 2.88\% of the time, users give up without creating a formula that compiles and 7.48\% of the time, the user makes 5 or more edits to correct an error. Upon investigating some of these longer invalid subpaths, we found Alloy's error reporting to be pretty effective for syntax errors but ambiguous at best, and misleading at worst, for type errors. 

Alloy formulas either produce sets or booleans. Oftentimes, a novice's type mistake is producing a set when a boolean is expected, or vice versa. However, Alloy's type-checking error reports often fail to make this clear. %Moreover, the report can even be misleading. 
To demonstrate, consider the following chain of edits for a \CodeIn{classroom\_fol} \CodeIn{inv15} attempt in which the user starts with a subformula that evaluates to a boolean when a set is expected:

\begin{footnotesize}
\begin{Verbatim}[frame=lines,rulecolor=\color{lightgray}]
inv15 T: \Blue{all} p:Person | {\setlength{\fboxsep}{1pt}\colorbox{red!30}{\Blue{some}}} (Teacher \Blue{in} p.^Tutors)
inv15 T: \Blue{all} p:Person | ({\setlength{\fboxsep}{1pt}\colorbox{red!30}{\Blue{some} Teacher}}) \Blue{in} p.^Tutors
inv15 T: \Blue{all} p:Person | {\setlength{\fboxsep}{1pt}\colorbox{red!30}{\Blue{some} Teacher}} \Blue{in} (p.^Tutors)
\end{Verbatim}
\end{footnotesize}

%inv15 S: \Blue{all} p:Person | (\Blue{some} Teacher) \Blue{in} p.^Tutors\colorbox{red!30}{)}
%inv15 S: \Blue{all} p:Person | (\Blue{some} Teacher \Blue{in} (p.^Tutors) \colorbox{red!30}{\}}

\noindent The highlights show the guidance given by Alloy's error report. For line 1, the user is only told ``\textit{the formula fails to typecheck}'' with the unary operator \CodeIn{some} the only highlight. This report fails to inform the user that ``\CodeIn{Teacher in p.Tutors}'' evaluates to a boolean and ``\CodeIn{some}'' expects a set, % to reason over whether that set is empty, 
leading to the type issue. For lines 2 and 3, the text of the error report --  ``\textit{This must be a set or relation. Instead, it has the following possible type(s): PrimitiveBoolean}.'' -- is more robust. However, 
%this report message is based on the idea that the core formula is supposed to be ``\CodeIn{a in b},'' even though the user could be trying to write ``\CodeIn{some a}.'' Moreover, 
the highlighted formula ``\CodeIn{some Teacher}'' is actually a valid type checked formula, which introduces confusion.
As this chain of edits shows, the user did not pick up on the core of the type issue, as the user kept trying to solve the problem by inserting parenthesis but only changed the type error from ``\CodeIn{some \textbf{[boolean]}}'' to ``\CodeIn{\textbf{[boolean]} in [set]}.''
%As this chain of edits shows, the user did not pick up on the core of the type issue, as the user only tried to insert parenthesis to change the scope without restructuring the formula. %, but the main problem, expressing \CodeIn{[some] [boolean]}, never gets removed. 

%Another frequent type of mistake is that users would often write a formula that evaluates to a set when the formula needs to evaluate to a boolean. However, 
The error report is equally ambiguous when a user writes a formula that evaluates to a set when a boolean is expected. To illustrate, consider the following attempt for \CodeIn{courses\_v1} \CodeIn{inv12}:

\begin{footnotesize}
\begin{Verbatim}[frame=lines,rulecolor=\color{lightgray}]
inv12 T: \Blue{all} s:Student | {\setlength{\fboxsep}{1pt}\colorbox{red!30}{s.enrolled.grades}}
inv12 T: \Blue{all} s:Student | {\setlength{\fboxsep}{1pt}\colorbox{red!30}{s.enrolled.grades.s}}
inv12 T: \Blue{all} s:Student | {\setlength{\fboxsep}{1pt}\colorbox{red!30}{s.(s.enrolled.grades)}}
\end{Verbatim}
\end{footnotesize}

%inv12 T: \Blue{all} s:Student, c:Course | \colorbox{red!30}{c.grades}
%inv12 T: \Blue{all} s:Student, c:Course | \colorbox{red!30}{s.(c.grades)}

With each edit, the user is informed that the highlighted text must ``\textit{be a formula expression.},'' There is no definition of a ``formula expression'' and no mention of the expectation that this formula needs to evaluate to a boolean. As a result, this user keeps trying to edit the highlighted formula as if the type error is within the formula. However, the user fails to realize that the edited formulas typecheck individually, but the quantified formula encompassing these formulas expects a boolean subformula not a set subformula. % seems to have misunderstood and thought there was a type error in the highlighted formula itself, when in fact, these highlighted formulas are all valid. The issue that that they produce a set of atoms and need to be used as a subformula within some relational logic statement, such as ``\CodeIn{some s.enrolled.grades},'' since the quantified formula expects to check whether each atom in the domain is true or false over a boolean subformula. 

\begin{impact}
\vspace{-.5ex}
Alloy's error reporting for types is too in the weeds and fails to convey when there is a highlevel mismatch between sets and booleans as ``types.'' Error reports and educational efforts should focus on highlighting how different classifications of operators require one or the other.
\vspace{-.5ex}
%Alloy's error reporting for types is essentially too in the weeds and fails to convey when there is a highlevel mismatch between sets and boolean producing formulas.
%While the error reporting can often be adequate, we have found that we found Alloy's type checking error reporting could be misleading. 
%Error reports and educational efforts should focus on highlighting if a formula produces a set or a boolean, and how different classifications of operators require one or the other.
\end{impact}
\subsection{RQ4: How do novice users' correct answers differ from the oracle formula?}

Oftentimes, the oracle solution represents one of several ways in which the property could be expressed, as seen in Tables~\ref{tab:unique_type_detailed} and \ref{tab:semantic} in which the number of syntactically unique correct solutions is larger than the number of semantically unique correct solutions. Therefore, we wanted to explore the ways novice users' 
%commonly write correct properties to gain insights from ways in which a novice user's 
thinking may differ from the expert-written oracles. 

First, we explore how frequently the oracle solution appears as a correct submission. Table~\ref{tab:oracle_loc} displays the location of the oracle submission derived by ranking all syntactically unique correct submissions in decreasing order from most to least duplicate submissions. Then, we looked for the location of the oracle submission within that list. Column \textbf{Type} displays the type of logical exercises under consideration based on the oracle formula. Column \textbf{Top 1} shows the number of exercises for that type in which the oracle is first in the ranked list and associated Column \textbf{Rate} is what percentage this number is out of the total number of exercises of that type.  Table~\ref{tab:oracle_loc} also displays information for whether the oracle is in the \textbf{Top 5}, \textbf{Top 10}, or never submitted (\textbf{Not Sub}).

\begin{table}[b]
  \centering
  \vspace{1ex}
\caption{Location of Oracle Among Correct Subs}
  \label{tab:oracle_loc}
  \resizebox{!}{.1\columnwidth}{
  \begin{tabular}{l|rr|rr|rr|rr}\hline
\rowcolor[HTML]{C0C0C0} 
\multicolumn{1}{c|}{\cellcolor[HTML]{C0C0C0}\textbf{Type}} & \multicolumn{1}{c}{\cellcolor[HTML]{C0C0C0}\textbf{Top 1}} & \multicolumn{1}{c|}{\cellcolor[HTML]{C0C0C0}\textbf{Rate}} & \multicolumn{1}{c}{\cellcolor[HTML]{C0C0C0}\textbf{Top 5}} & \multicolumn{1}{c|}{\cellcolor[HTML]{C0C0C0}\textbf{Rate}} & \multicolumn{1}{c}{\cellcolor[HTML]{C0C0C0}\textbf{Top 10}} & \multicolumn{1}{c|}{\cellcolor[HTML]{C0C0C0}\textbf{Rate}} & \multicolumn{1}{c}{\cellcolor[HTML]{C0C0C0}\textbf{Not Sub}} & \multicolumn{1}{c}{\cellcolor[HTML]{C0C0C0}\textbf{Rate}} \\\hline
\textbf{RL} & 22 & 43.14 & 44 & 86.27 & 49 & 96.08 & 3 & 5.88 \\
\rowcolor[HTML]{EFEFEF} 
\textbf{PL} & 1 & 100.00 & 1 & 100.00 & 1 & 100.00 & 0 & 0.00 \\
\textbf{FOL} & 25 & 26.32 & 58 & 61.05 & 65 & 68.42 & 24 & 25.26 \\
\rowcolor[HTML]{EFEFEF} 
\textbf{LTL} & 12 & 33.33 & 21 & 58.33 & 22 & 61.11 & 9 & 25.00 \\ \hline
\rowcolor[HTML]{DAE8FC}  
\textbf{SUM/AVG} & \textbf{60} & \textbf{32.79} & \textbf{156} & \textbf{85.25} & \textbf{169} & \textbf{92.35} & \textbf{36} & \textbf{19.67} \\ \hline 
\end{tabular}
}
\end{table}

For RL exercises, the oracle is the top submission 43.14\% of the time and in the top 10 96.08\% of the time. Rarely (5.88\%) the oracle is never submitted. The oracle is only the top submission for 26.32\% of FOL exercises and 33.33\% of LTL exercises. By location 10, the oracle is only present in 68.42\% and 61.11\% of the exercises respectively. Moreover, it is common for the oracle to never be submitted, which happens about 25\% of the time for both FOL and LTL exercises. 
%Overall, FOL and LTL exercises are both more complex, but also more open-ended. For instance, for FOL exercises, depending on what the user selects as the domain, the formulas can vary widely. 
RL exercise use the smallest subset of operators, and this seems to translate to less diversity in how a property can be correctly expressed. This is further supported by Table~\ref{tab:unique_type_detailed}, in which 15.7\% of correct RL submissions are syntactically unique but 21.85\% of FOL submissions are syntactically unique. Interestingly, only 12.58\% of LTL exercises are syntactically unique, which indicates that users frequently write the same correct submissions, but fail to utilize the same structure as the oracle.

\begin{table}
  \centering
  \caption{Differences - Oracle and Hottest Hit Corr. Subs}
  \label{tab:diff_oracle_sub}
  \resizebox{!}{.29\columnwidth}{
  \begin{tabular}{lrr}\hline
\rowcolor[HTML]{C0C0C0} 
\multicolumn{1}{l}{\cellcolor[HTML]{C0C0C0}\textbf{Difference}} & \multicolumn{1}{c}{\cellcolor[HTML]{C0C0C0}\textbf{\# Occ}} & \multicolumn{1}{c}{\cellcolor[HTML]{C0C0C0}\textbf{Rate}} \\ \hline
\textbf{Upscaled to quantification} & 21 & 11.17 \\
\rowcolor[HTML]{EFEFEF} 
\textbf{Upscaled to nested quantification} & 32 & 17.02 \\
\textbf{Upscaled to further nested quantification} & 2 & 1.06 \\
\rowcolor[HTML]{EFEFEF} 
\textbf{Downscaled removed nested quantification} & 11 & 5.85 \\
\textbf{Downscaled removed quantification} & 10 & 5.32 \\
\rowcolor[HTML]{EFEFEF} 
\textbf{Different quantification domains} & 11 & 5.85 \\
\textbf{Similar but different order of operands} & 12 & 6.38 \\
\rowcolor[HTML]{EFEFEF} 
\textbf{Similar but formed different sets} & 7 & 3.72 \\
\textbf{Similar but different operator} & 19 & 10.11 \\
\rowcolor[HTML]{EFEFEF} 
\textbf{Similar but additional trivial operator} & 1 & 0.53 \\
\textbf{Expanded out formula} & 2 & 1.06 \\
\rowcolor[HTML]{EFEFEF} 
\textbf{No Change} & 60 & 31.91 \\ \hline
\rowcolor[HTML]{DAE8FC}  
\textbf{SUM/AVG} & \textbf{188} & \textbf{100.00} \\ \hline
\end{tabular}}
\end{table}

Second, to build a better understanding of why LTL exercises often do not contain the oracle submission, and to further investigate the differences users make compared to the oracle, we manually investigated the differences between the top 1 correct solution and the oracle. Table~\ref{tab:diff_oracle_sub} displays for each identified difference the number of top 1 correct submissions that have this difference (column \textbf{\# Occ}) and the percentage that this accounts for out of all top 1 submissions (column \textbf{Rate}). 7 exercises had no correct submission and 12 exercises had 2 main differences, which gives us a total number (column \textbf{\# Occ}) of 188.

Nearly one-third of the top 1 correct submissions are syntactically equivalent to the oracle. For the remaining two-thirds, 35 submissions are similar but involve minor differences, such swapping the order of operands for commutative operators. Many of these submissions are LTL exercises in which the location of the temporal operator \CodeIn{always} varies. In addition, the three of the four ``similar but additional trivial operator'' are LTL exercises in which the user appends an unnecessary leading \CodeIn{always} operator. Together, these similar submissions highlight common modeling preferences that result in the oracle being less present for LTL exercises.

\begin{impact}
Novice users prefer leading with temporal operators, while the oracle solutions insert them close to the subformula of interest. For instance, a novice user will write ``\CodeIn{always all a: A $\vert$ F}'' while the oracle will use  ``\CodeIn{all a : A $\vert$ always F}.''
\end{impact}

Overall, the most prevalent difference is that novice users increase the degree of quantification in the formula, which is seen in 55 (30.5\%) of the top 1 solutions. This indicates that novice users often prefer to try and explicitly outline relationships between atoms or sequences of events. Given this preference, we wanted to see to what degree novice users write more verbose formulas. 

Table~\ref{tab:ast_diff} displays the Abstract Syntax Tree (AST) difference between the oracle solution and all of the correct submissions. Column \textbf{Type} conveys the type of logic, with a final row that reflects all correct submission. Column \textbf{Dif$_{Tot}$} displays the average difference in total number of AST nodes between the oracle and all correct submissions. The next four columns (\textbf{Dif$_{RL}$}, \textbf{Dif$_{PL}$}, \textbf{Dif$_{FOL}$}, \textbf{Dif$_{LTL}$}) displays the average difference in total number of AST nodes of each type of logic (RL, PL, FOL, and LTL) between the oracle and all correct submissions. For these columns, a negative number means the oracle had less AST nodes. To give context to how significant the difference in nodes is, column \textbf{Dif$_{Mag}$} displays average magnitude in difference between the oracle and correct submissions. 

Overall, Table~\ref{tab:ast_diff} reinforces the implication from Table~\ref{tab:diff_oracle_sub} that novice users often write larger formulas. RL and PL exercises see the largest increase in size, which is expected, as one of the common reasons the formula is larger based on Table~\ref{tab:diff_oracle_sub} is that the user is adding at least one new quantified formula to their solution compared to the oracle. Not surprisingly, this difference in the number of AST nodes translates to the oracle being on average 0.38$\times$ smaller for both RL and PL exercises. However, even FOL exercises, which already have a quantified formula, also see a large increase in the number of AST nodes, with the oracle on average being 8.17 nodes smaller. Table~\ref{tab:diff_oracle_sub} highlights that users even up-scaled existing quantified formulas to have nested or even further nested quantification. For LTL exercises, while the novice correct submissions are still on average larger, the difference is only 3.7 nodes. However, these 3.7 nodes result in an oracle that is on average 0.77$\times$ the size of the correct submission, which is similar to FOL exercises in which the 8.17 nodes means the oracle is on average 0.81$\times$ the size of the correct submissions.

%\begin{table*}
%\centering
%\caption{AST and Operator Differences Between Oracle and Submissions}
%\begin{footnotesize}
%\input{tables/ast_diff}
%\end{footnotesize}
%\label{tab:common_mistakes}
%\end{table*}

\begin{table}
  \centering
\caption{AST Differences Between Oracle and Corr. Subs}
  \label{tab:ast_diff}
  \resizebox{!}{.12\columnwidth}{
  \begin{tabular}{l|rrrrrr}\hline
\rowcolor[HTML]{C0C0C0} 
 & \multicolumn{1}{c}{\cellcolor[HTML]{C0C0C0}\textbf{Dif$_{Tot}$}} & \multicolumn{1}{c}{\cellcolor[HTML]{C0C0C0}\textbf{Dif$_{RL}$}} & \multicolumn{1}{c}{\cellcolor[HTML]{C0C0C0}\textbf{Dif$_{PL}$}} & \multicolumn{1}{l}{\cellcolor[HTML]{C0C0C0}\textbf{Dif$_{FOL}$}} & \multicolumn{1}{c}{\cellcolor[HTML]{C0C0C0}\textbf{Dif$_{LTL}$}} & \multicolumn{1}{c}{\cellcolor[HTML]{C0C0C0}\textbf{Dif$_{Mag}$}} \\ \hline
\textbf{RL} & -11.53 & -2.49 & -0.69 & -1.16 & -0.02 & 0.38 \\
\rowcolor[HTML]{EFEFEF} 
\textbf{PL} & -10.74 & -1.79 & -0.03 & -1.26 & 0.00 & 0.38 \\
\textbf{FOL} & -8.17 & -1.87 & -0.82 & -0.52 & 0.00 & 0.81 \\
\rowcolor[HTML]{EFEFEF} 
\textbf{LTL} & -3.70 & -0.89 & -0.43 & -0.24 & -0.07 & 0.77 \\ \hline
\rowcolor[HTML]{DAE8FC}  
\textbf{SUM/AVG} & \textbf{-9.08} & \textbf{-2.03} & \textbf{-0.76} & \textbf{-0.72} & \textbf{-0.01} & \textbf{0.66} \\ \hline
\end{tabular}}
\end{table}

\begin{impact}
Educators and development toolsets can focus on how to refactor verbose formulas into more condense versions that can be (1) more readable and/or (2) more efficient to analyze. 
%When novice modelers write a property correctly, they are likely to write longer formulas that more explicitly outline relationships (RL, PL, FOL) or steps (LTL).
\end{impact}

\subsection{RQ5: How often do novice users reach a correct answer?}
Since only a forth of all submissions are correct, there is a high likelihood that some attempts end without reaching a correct answer. To explore this, for each model, Table~\ref{tab:correct_first} displays what percentage of attempts have a correct submission, what percentage of attempts have a correct answer as the first submission, what percentage of attempts never produce a correct answer, and for those without a correct answer, the average number of submissions made until the user gave up. Since the user may switch between different types of logic within each attempt, we do not report type-based information.

%Considering that 25.4\% of the submissions are correct, we were curious to explore how novice users arrive at the correct answer. For each model, Table~\ref{tab:correct_first} displays what percentage of attempts have a correct submission, what percentage of attempts have a correct answer as the first submission, what percentage of attempts never produce a correct answer, and for those without a correct answer, the average number of submissions made until the user gave up. Since the user may switch between different types of logic within each attempt, we do not report any type-based information.

Across all attempts, 67.38\% of the time the user gets to a correct answer and notably users write a correct answer first in 31.23\% of all attempts. The complexity of the underlying base model, as captured in Table~\ref{tab:model_stats}, does have some roll in users reaching a correct submission, as the trash models and graph models have some of the highest rate of attempts with a correct submission. However, the complexity of the underlying base model does not necessarily translate to how often novices get solutions correct on the first attempt. For instance, while \CodeIn{trash\_fol} has the highest percentage and is a simple model, \CodeIn{classroom\_fol} has the second highest percentage and is notable more complex. Instead, correct on first attempts were more closely connected to exercises that occurred earlier in a model. Most models are set up such that the difficulty of the exercise increases as you progress through the exercises.

\begin{table}
\centering
\caption{Details for Correct on First Attempt}
\resizebox{!}{.35\columnwidth}{
\begin{tabular}{lrrrr} \hline
\rowcolor[HTML]{C0C0C0} 
\multicolumn{1}{c}{\cellcolor[HTML]{C0C0C0}\textbf{Model}} & \multicolumn{1}{c}{\cellcolor[HTML]{C0C0C0}\textbf{\begin{tabular}[c]{@{}c@{}}\% Atmps \\ with Corr\end{tabular}}} & \multicolumn{1}{c}{\cellcolor[HTML]{C0C0C0}\textbf{\% First Try}} & \multicolumn{1}{c}{\cellcolor[HTML]{C0C0C0}\textbf{\begin{tabular}[c]{@{}c@{}}\% Atmpts \\ Never Corr\end{tabular}}} & \multicolumn{1}{c}{\cellcolor[HTML]{C0C0C0}\textbf{\begin{tabular}[c]{@{}c@{}}Avg steps\\ till give up\end{tabular}}} \\ \hline
\textbf{classroom\_fol} & 84.26 & 51.47 & 15.74 & 6.44 \\
\rowcolor[HTML]{EFEFEF} 
\textbf{classroom\_rl} & 78.03 & 21.97 & 21.97 & 5.14 \\
\textbf{courses\_v1} & 48.53 & 21.13 & 51.47 & 6.07 \\
\rowcolor[HTML]{EFEFEF} 
\textbf{courses\_v2} & 47.87 & 23.89 & 52.13 & 5.01 \\
\textbf{cv\_v1} & 45.85 & 23.32 & 54.15 & 5.83 \\
\rowcolor[HTML]{EFEFEF} 
\textbf{cv\_v2} & 65.22 & 36.23 & 34.78 & 8.75 \\
\textbf{graphs} & 74.14 & 41.88 & 25.86 & 3.72 \\
\rowcolor[HTML]{EFEFEF} 
\textbf{lts} & 58.91 & 32.47 & 41.09 & 6.47 \\
\textbf{productionLine\_v1} & 69.23 & 29.06 & 30.77 & 3.18 \\
\rowcolor[HTML]{EFEFEF} 
\textbf{productionLive\_v2} & 73.94 & 30.44 & 26.06 & 4.40 \\
\textbf{productionLine\_v3} & 75.38 & 25.79 & 24.62 & 5.26 \\
\rowcolor[HTML]{EFEFEF} 
\textbf{socialMedia} & 68.08 & 28.05 & 31.92 & 6.64 \\
\textbf{trainStation\_fol} & 77.29 & 23.69 & 22.71 & 9.19 \\
\rowcolor[HTML]{EFEFEF} 
\textbf{trianStation\_ltl} & 60.91 & 22.73 & 39.09 & 8.01 \\
\textbf{trash\_fol} & 86.20 & 61.78 & 13.80 & 3.76 \\
\rowcolor[HTML]{EFEFEF} 
\textbf{trash\_ltl} & 80.59 & 40.75 & 19.41 & 8.03 \\
\textbf{trash\_rl} & 76.35 & 44.49 & 23.65 & 2.38 \\ \hline
\rowcolor[HTML]{DAE8FC} 
\textbf{SUM/AVG} & \textbf{67.38} & \textbf{31.23} & \textbf{32.62} & \textbf{5.85} \\ \hline
\end{tabular}
}
\label{tab:correct_first}
\end{table}

\begin{table*}
\centering
\caption{Common Mistakes Made by Novices (out of submissions that at least 10 novice users made)}
\resizebox{!}{.38\columnwidth}{
\begin{tabular}{l|rrrrrr||rrrrrr||rr}\hline
\rowcolor[HTML]{C0C0C0} 
\multicolumn{1}{c|}{\cellcolor[HTML]{C0C0C0}} & \multicolumn{2}{c}{\cellcolor[HTML]{C0C0C0}\textbf{Both}} & \multicolumn{2}{c}{\cellcolor[HTML]{C0C0C0}\textbf{Over}} & \multicolumn{2}{c||}{\cellcolor[HTML]{C0C0C0}\textbf{Under}} & \multicolumn{2}{c}{\cellcolor[HTML]{C0C0C0}\textbf{RL}} & \multicolumn{2}{c}{\cellcolor[HTML]{C0C0C0}\textbf{FOL}} & \multicolumn{2}{c||}{\cellcolor[HTML]{C0C0C0}\textbf{LTL}} & \multicolumn{2}{c}{\cellcolor[HTML]{C0C0C0}\textbf{Total}} \\ 
\rowcolor[HTML]{C0C0C0} 
\multicolumn{1}{c|}{\multirow{-2}{*}{\cellcolor[HTML]{C0C0C0}\textbf{Difference}}} & \multicolumn{1}{c}{\cellcolor[HTML]{C0C0C0}\textbf{\#Occ}} & \multicolumn{1}{c}{\cellcolor[HTML]{C0C0C0}\textbf{\#Sub}} & \multicolumn{1}{c}{\cellcolor[HTML]{C0C0C0}\textbf{\#Occ}} & \multicolumn{1}{c}{\cellcolor[HTML]{C0C0C0}\textbf{\#Sub}} & \multicolumn{1}{c}{\cellcolor[HTML]{C0C0C0}\textbf{\#Occ}} & \multicolumn{1}{c||}{\cellcolor[HTML]{C0C0C0}\textbf{\#Sub}} & \multicolumn{1}{c}{\cellcolor[HTML]{C0C0C0}\textbf{\#Occ}} & \multicolumn{1}{c}{\cellcolor[HTML]{C0C0C0}\textbf{\#Sub}} & \multicolumn{1}{c}{\cellcolor[HTML]{C0C0C0}\textbf{\#Occ}} & \multicolumn{1}{c}{\cellcolor[HTML]{C0C0C0}\textbf{\#Sub}} & \multicolumn{1}{c}{\cellcolor[HTML]{C0C0C0}\textbf{\#Occ}} & \multicolumn{1}{c||}{\cellcolor[HTML]{C0C0C0}\textbf{\#Sub}} & \multicolumn{1}{c}{\cellcolor[HTML]{C0C0C0}\textbf{\#Occ}} & \multicolumn{1}{c}{\cellcolor[HTML]{C0C0C0}\textbf{\#Sub}} \\ \hline
\textit{\textbf{More   complex quant. domain}} & 1 & 11 & 0 & 0 & 3 & 53 & 0 & 0 & 4 & 64 & 0 & 0 & 4 & 64 \\
\rowcolor[HTML]{EFEFEF} 
\textit{\textbf{Narrowed   quant. domain}} & {\color[HTML]{000000} 16} & {\color[HTML]{000000} 224} & {\color[HTML]{000000} 4} & {\color[HTML]{000000} 160} & {\color[HTML]{009901} \textbf{44}} & {\color[HTML]{009901} \textbf{846}} & 0 & 0 & {\color[HTML]{009901} \textbf{64}} & {\color[HTML]{009901} \textbf{1230}} & 0 & 0 & {\color[HTML]{009901} \textbf{64}} & {\color[HTML]{009901} \textbf{1230}} \\
\textit{\textbf{Different   quant. domain}} & 1 & 34 & 0 & 0 & 5 & 149 & 0 & 0 & 6 & 183 & 0 & 0 & 6 & 183 \\
\rowcolor[HTML]{EFEFEF} 
\textit{\textbf{Simplified   quant. Domain}} & 6 & 108 & 0 & 0 & 0 & 0 & 0 & 0 & 6 & 108 & 0 & 0 & 6 & 108 \\
\textit{\textbf{Upscaled   quant. level trying to state relationships}} & {\color[HTML]{009901} \textbf{39}} & {\color[HTML]{009901} \textbf{781}} & {\color[HTML]{009901} \textbf{17}} & {\color[HTML]{009901} \textbf{387}} & {\color[HTML]{009901} \textbf{53}} & {\color[HTML]{009901} \textbf{981}} & 0 & 0 & {\color[HTML]{009901} \textbf{101}} & {\color[HTML]{009901} \textbf{1990}} & {\color[HTML]{009901} \textbf{8}} & {\color[HTML]{009901} \textbf{159}} & {\color[HTML]{009901} \textbf{109}} & {\color[HTML]{009901} \textbf{2149}} \\
\rowcolor[HTML]{EFEFEF} 
\textit{\textbf{Incorrect   order of nested quant.}} & 3 & 42 & 1 & 36 & 0 & 0 & 0 & 0 & 4 & 78 & 0 & 0 & 4 & 78 \\
\textit{\textbf{Nested   quantification disjoint mistake}} & 13 & 322 & 0 & 0 & 4 & 91 & 0 & 0 & 17 & 413 & 0 & 0 & 17 & 413 \\
\rowcolor[HTML]{EFEFEF} 
\textit{\textbf{Downscaled   quant. level leading to incorr. expr range}} & {\color[HTML]{009901} \textbf{17}} & {\color[HTML]{009901} \textbf{241}} & {\color[HTML]{009901} \textbf{7}} & {\color[HTML]{009901} \textbf{137}} & {\color[HTML]{000000} 10} & {\color[HTML]{000000} 216} & {\color[HTML]{009901} \textbf{18}} & {\color[HTML]{009901} \textbf{309}} & 14 & 251 & 2 & 34 & 34 & 594 \\
\textit{\textbf{Downsacled   quant. Level leading to new subformula}} & 2 & 27 & 3 & 43 & 1 & 12 & 5 & 69 & 1 & 13 & 0 & 0 & 6 & 82 \\
\rowcolor[HTML]{EFEFEF} 
\textit{\textbf{Tried to   inverted concept}} & 2 & 25 & 4 & 70 & 4 & 124 & 2 & 37 & 6 & 144 & 2 & 38 & 10 & 219 \\
\textit{\textbf{Trying to   explictly outline steps}} & 1 & 12 & 0 & 0 & 1 & 12 & 0 & 0 & 0 & 0 & 2 & 24 & 2 & 24 \\
\rowcolor[HTML]{EFEFEF} 
\textit{\textbf{Incorrect use   of operator}} & {\color[HTML]{009901} \textbf{59}} & {\color[HTML]{009901} \textbf{1074}} & {\color[HTML]{009901} \textbf{36}} & {\color[HTML]{009901} \textbf{616}} & {\color[HTML]{009901} \textbf{61}} & {\color[HTML]{009901} \textbf{1194}} & {\color[HTML]{009901} \textbf{14}} & {\color[HTML]{009901} \textbf{238}} & {\color[HTML]{009901} \textbf{130}} & {\color[HTML]{009901} \textbf{2438}} & {\color[HTML]{009901} \textbf{12}} & {\color[HTML]{009901} \textbf{208}} & {\color[HTML]{009901} \textbf{156}} & {\color[HTML]{009901} \textbf{2884}} \\
\textit{\textbf{Operator not   commumtative}} & 7 & 105 & 7 & 121 & 6 & 153 & 1 & 14 & 16 & 294 & 3 & 71 & 20 & 379 \\
\rowcolor[HTML]{EFEFEF} 
\textit{\textbf{Missing   operator}} & 1 & 10 & 4 & 71 & 11 & 250 & 1 & 10 & 7 & 175 & {\color[HTML]{009901} \textbf{8}} & {\color[HTML]{009901} \textbf{146}} & 16 & 331 \\
\textit{\textbf{Incorrectly   scoped expr, missing parentheses}} & 1 & 23 & 0 & 0 & 0 & 0 & 0 & 0 & 1 & 23 & 0 & 0 & 1 & 23 \\
\rowcolor[HTML]{EFEFEF} 
\textit{\textbf{Incorrect   application of extension signature}} & {\color[HTML]{009901} \textbf{16}} & {\color[HTML]{009901} \textbf{235}} & {\color[HTML]{009901} \textbf{7}} & {\color[HTML]{009901} \textbf{165}} & {\color[HTML]{009901} \textbf{46}} & {\color[HTML]{009901} \textbf{850}} & 3 & 46 & {\color[HTML]{009901} \textbf{65}} & {\color[HTML]{009901} \textbf{1181}} & 1 & 23 & {\color[HTML]{009901} \textbf{69}} & {\color[HTML]{009901} \textbf{1250}} \\
\textit{\textbf{Subportion of   total concept}} & 3 & 74 & 6 & 143 & 22 & 449 & {\color[HTML]{009901} \textbf{5}} & {\color[HTML]{009901} \textbf{128}} & 23 & 492 & 3 & 46 & 31 & 666 \\
\rowcolor[HTML]{EFEFEF} 
\textit{\textbf{Wrong   understanding of exercise}} & 9 & 156 & 1 & 13 & 0 & 0 & {\color[HTML]{009901} \textbf{4}} & {\color[HTML]{009901} \textbf{61}} & 3 & 55 & {\color[HTML]{009901} \textbf{3}} & {\color[HTML]{009901} \textbf{53}} & 10 & 169 \\  \hline
\rowcolor[HTML]{DAE8FC} 
\textbf{Total} & \textbf{197} & \textbf{3504} & \textbf{97} & \textbf{1962} & \textbf{271} & \textbf{5380} & \textbf{53} & \textbf{912} & \textbf{468} & \textbf{9132} & \textbf{44} & \textbf{802} & \textbf{565} & \textbf{10846} \\  \hline
\end{tabular}
}
\label{tab:common_mistakes}
\end{table*}

Novice users struggle with \CodeIn{courses\_v1}, \CodeIn{courses\_v2} and \CodeIn{cv\_v1}, in which less than 50\% of all attempts made for these models have a correct submission. Both \CodeIn{courses\_v1} and \CodeIn{courses\_v2} have two signatures that contain a relation with the same name, which adds additional complexity to the exercises as users often have to properly scope formulas to remove ambiguity in which relation is getting accessed. For \CodeIn{cv\_v1} and additionally three other models low correctness rates -- \CodeIn{cv\_v2}, \CodeIn{socialMedia} and \CodeIn{trainStation\_ltl} -- all of these models require first order logic to solve most of their exercises. Furthermore,  \CodeIn{socialMedia} and \CodeIn{trainStation\_ltl} have the largest oracle submissions in terms of number of AST nodes, indicating that the properties-to-be-expressed for these models are more complex compared to the other models. Complexity of the exercise likely influences more than just how often users reach a correct solution. Across all models, there is a dropoff in the number of attempts for the later exercises for a model. This indicates that users tend to stop working on a model as the exercises get harder. 

\begin{impact}
A third of the time, novice users give up, preventing them from fully learning the intended lesson. Since novices give up quickly ($<$6 tries), educational tools should provide more guidance or positive reinforcement to keep students engaged.
\vspace{-1.5ex}
\end{impact}

\subsection{RQ6: How often do novice users make the same logical mistake?}

As indicated by Table~\ref{tab:unique_type_detailed}, novice users often repeat the same mistake. Understanding these common mistakes can help researchers focusing on fault localization and repair, as well as educators teaching formal methods. To give further guidance on this, we explore a subset of ``hot hit'' submissions in which the same syntactic submission is made at least 10 times. We manually investigated each ``hot hit'' and label it with core tenants of the mistake. Table~\ref{tab:common_mistakes} displays the common mistakes made broken down by classification of submission and type of logic. Column \textbf{\#Occ} represent the number of unique ``hot hit'' submissions of that mistake and column \textbf{\#Sub} represents how many total number of all ``hot hit'' submissions of that mistake. In green, we highlight the top 4 mistakes.

Across all logic types, the single most common mistake is the incorrect use of an operator. Our dataset includes a detailed breakdown of which specific operators are incorrectly applied and at what rates. The most common misapplication is related to the use of the subset, relational join, and implication operators. For subset mistakes, a common but very subtle mistake is for novice users to use subset in place of set intersection. For their formulas, these two operators often behave the same except for when the left-hand side of the operator is empty. For example, consider the following submission (s) to \CodeIn{inv6} from the \CodeIn{productionLine\_v2} model that 20 novice users submitted (19.61\% of inv7's both over- and underconstrained submissions) and the closest correct solution (c):

\begin{footnotesize}
\begin{Verbatim}[frame=lines,rulecolor=\color{lightgray}]
\Green{// Components built of dangerous parts are also dangerous}
s: \Blue{all} c: Component | c.parts \Blue{in} Dangerous => c \Blue{in} Dangerous
c: \Blue{all} c: Component | \Blue{some} c.parts & Dangerous => c \Blue{in} Dangerous
\end{Verbatim}
\end{footnotesize}

In this case, if \CodeIn{c.parts} is empty, then \CodeIn{c.parts in Dangerous} evaluates to \CodeIn{true}, which prevents scenarios from being generated in which there are no dangerous components. This is a subtle mistake that is also common in the misuse of the subset exclusion operator, where users would express \CodeIn{a !in b} instead of \CodeIn{no a \& b}. In addition, users often made the wrong choice when selecting between two closely related operators, selecting subset (`\CodeIn{in}') instead of set equality (`\CodeIn{=}'), transitive closure (`\CodeIn{\ACaret{}}') instead of reflexive transitive closure (`\CodeIn{\AStar{}}') and implication (`\CodeIn{=>}') instead of biconditional (`\CodeIn{<=>}').

\begin{impact}
Educators should focus on the subtle difference between similar but different operators, including combinations of operators that together behave similarly. In addition, educators should illustrate how logical operators behave in corner case settings, such as when reasoning over the empty set. 
\end{impact}

%With FOL submissions prevalent in Table~\ref{tab:common_mistakes}, there are fundamental misunderstandings novice users have about the appropriate application of FOL. Half of the frequently repeated classification of mistakes (column \textbf{Mistakes}) involve novices introducing mistakes related to quantification, which reiterates our observations in RQ1 that FOL formulas gave novice users problems. 
FOL submissions are prevalent in Table~\ref{tab:common_mistakes}, accounting for 468 of the 565 hot hit submissions. In addition, half of the frequently repeated classification of mistakes (column \textbf{Mistakes}) involve novices introducing errors related to quantification. Therefore, there are clearly fundamental misunderstandings novice users have about how and when to use quantification.
The most common quantifier mistake, and the second most common overall mistake, is that novices try to upscale the level of quantification in an attempt to explicitly outline relationships, where upscale means the submission inserted more quantifiers than the oracle. 
%This issue occurred across 109 different hot hit submissions which together represent 2,149 submissions. 
As an example, consider the following submission for \CodeIn{classroom\_rl} \CodeIn{inv5}:

\begin{footnotesize}
\begin{Verbatim}[frame=lines,rulecolor=\color{lightgray}]
\Green{//There classes assigned to teachers.}
s: \Blue{all} t:Teacher| \Blue{some} t.Teaches
c: \Blue{some} Teacher.Teaches
\end{Verbatim}
\end{footnotesize}

that accounts for 30 (35.29\%) of the both over- and underconstrained submissions for \CodeIn{inv5}. To correct this formula, the user should exchange universal quantification  (`\CodeIn{all}') to existential quantification (`\CodeIn{some}').  While there are often ways to express a relational formula with an equivalent first-order logic formula, the increase in complexity of the formula structure introduces more opportunities for users to make a mistake. %Upscaling often creates underconstrained or both over- and underconstrained faults for FOL and LTL submissions. 

Hand in hand with upscaling, downscaling the level of quantification is a common mistake that often results in an expression being applied to the wrong set of elements. %Namely, at some point in the constraint, the novice user writes a relational join expression in which the left-hand side captures too broad of a set. Downscaling occurs in 40 ``hot hits'' that represent 676 user submissions. 
As an example, consider the submission for \CodeIn{inv1} for model \CodeIn{lts}:

\begin{footnotesize}
\begin{Verbatim}[frame=lines,rulecolor=\color{lightgray}]
\Green{//Each state has at least a transition.}
s: \Blue{some} State.trans
c: \Blue{all} s: State | \Blue{some} s.trans
\end{Verbatim}
\end{footnotesize}

that accounts for 28 (38.89\%) of the overconstrained submissions for \CodeIn{inv1}. Here, the user effectively writes the same expression \CodeIn{[x]\textbf{.trans}}; however, the expression is too broadly applied to the set of all \CodeIn{State} atoms, instead of checked for each individual \CodeIn{State} atom.  Downscaling creates both, over and underconstrained faults. This is expected, as our example highlights how downscaling can result in a multiplicity relationship between atoms being viewed as ``satisfied'' because at least one atom of the same type satisfies the relationship, which depending on the constraint, could make the model more or less permissive. 

%% removed - feels like a concluding statement that would be an "impact"
%Whether upscaling or downscaling, the user is often trying to express some relationship between atoms or sets of atoms. Novice users frequently make mistakes when trying to determine what level of quantification to use to express these relationships. 

\begin{impact}
Educators should focus on best practices for writing constraints that establish relationships between atoms and emphasize when the use of quantification is appropriate to express the relationship \textit{per atom} versus when not to use quantification to express the relationship per \textit{type} of the atom.
\end{impact}

The fourth most common overall mistake is for users to make the domain of a quantified formula too narrow, which frequently produces an underconstrained submission. A narrower quantification domain almost always occurred on models that had extension signatures and the novice user used an extension signature for the domain in place of the base signature. As an example, consider the following submission for \CodeIn{inv3} for model \CodeIn{courses\_v2} which accounts for 65 (40.37\%) of the underconstrained submissions:

\begin{footnotesize}
\begin{Verbatim}[frame=lines,rulecolor=\color{lightgray}]
\Green{// Courses must have teachers.}
s: \Blue{all} c : Course | \Blue{some} p : Professor | c \Blue{in} p.teaches
c: \Blue{all} c : Course | \Blue{some} p : Person | c \Blue{in} p.teaches
\end{Verbatim}
\end{footnotesize}

These narrowed quantification domains additionally account for the third most common mistake, the incorrect use of an extension signature. In addition to incorrectly using extension signatures in quantified domains, novice users also incorrectly used extension signatures within individual relational expressions. %, where an extension signature is used in conjunction with relational join to build sets when the parent signature is needed.

\begin{impact}
Educators should focus on when it is appropriate to use a parent versus an extension signature.
\end{impact}

Hand in hand with education, these guidelines are also blueprints for researchers to aid in improving debugging-oriented techniques, such as creating new repair templates or refining suspiciousness formulas for fault localization.
%In addition, these common mistakes provide clear blueprints for researchers to aid in improving debugging-oriented techniques.

%\begin{impact}
%Researchers should consider a small collection of higher-order mutants built around interchanging similar but different operators and changing the level of quantification of a formula to improve the suite of debugging tools for Alloy. For instance, for a faulty formula in the form ``\CodeIn{a !in b},'' a possible fix that should be investigated is the second order mutant: ``\CodeIn{no a \& b}.'' For the later, a faulty formula in the form ``\CodeIn{all a : A | some a.b},'' a possible fix that should be investigated is the second order mutant: ``\CodeIn{some A.b}.''
%\end{impact}

\begin{table}[b]
\vspace{1ex}
\centering
\caption{Frequency of Mutants Correcting Submissions}
\label{tab:osa}
\begin{footnotesize}
\begin{tabular}{l|rrr|rr} \hline
\rowcolor[HTML]{C0C0C0} 
\multicolumn{1}{c|}{\cellcolor[HTML]{C0C0C0}\textbf{Clx/Type}} & \multicolumn{1}{c}{\cellcolor[HTML]{C0C0C0}\textbf{\# Fixes}} & \multicolumn{1}{c}{\cellcolor[HTML]{C0C0C0}\textbf{\# Sub$_{f}$}} & \multicolumn{1}{c}{\cellcolor[HTML]{C0C0C0}\textbf{\% Sub$_{f}$}} & \multicolumn{1}{c}{\cellcolor[HTML]{C0C0C0}\textbf{\# Sub$_{m}$}} & \multicolumn{1}{c}{\cellcolor[HTML]{C0C0C0}\textbf{\% Sub$_{m}$}} \\ \hline
\textbf{Both} & 1089 & 882 & 6.24 & 173 & 1.22 \\
\rowcolor[HTML]{EFEFEF} 
\textbf{Over} & 1456 & 871 & 18.34 & 260 & 5.47 \\
\textbf{Under} & 1527 & 999 & 15.48 & 248 & 3.84 \\ \hline
\rowcolor[HTML]{EFEFEF} 
\textbf{RL} & 247 & 192 & 5.61 & 39 & 1.83 \\
\textbf{PL} & 47 & 29 & 11.69 & 9 & 2.34 \\
\rowcolor[HTML]{EFEFEF} 
\textbf{FOL} & 3092 & 2120 & 10.83 & 509 & 2.46 \\
\textbf{LTL} & 686 & 411 & 19.74 & 124 & 5.81 \\ \hline
\rowcolor[HTML]{DAE8FC} 
\textbf{SUM} & \textbf{4072} & \textbf{2752} & \textbf{10.86} & \textbf{681} & \textbf{29.57} \\ \hline
\end{tabular}
\end{footnotesize}
\end{table}

\subsection{RQ7: How often are users ``one'' mistake away?}
For incorrect submissions, we were curious to explore how often users were effectively ``one mistake away'' from a correct solution. To investigate this, we applied MuAlloy, a mutation testing tool~\cite{MuAlloyTemporal}, to all faulty submissions in the both, over and under categories. %Since the signature paragraphs for the models are static, we adapted MuAlloy to generate mutants for only the user-provided submission predicates. 
For each submission, we generated all first-order mutants. Then, if any of the mutated models are equivalent to the oracle submission we classify the submission as fixable and ``one mistake away.''  

We were interested in two viewpoints: (1) how often are submissions one mistake away and (2) what mutants correct faulty submissions. For the first viewpoint, Table~\ref{tab:osa} depicts the frequency at which mutants fix submissions broken down by the type of logic present in the submission and the classification of the submission. Column \textbf{\#Fixes} displays the total number of mutants that fixed a submission, column \textbf{\# Sub$_{f}$} shows the number of submissions that were fixed and column \textbf{\% Sub$_{f}$} shows what percentage of submissions \textbf{\# Sub$_{f}$} is. The number of fixes is more than the number of submissions fixed because some submissions could be fixed by multiple different mutations. To help explore this, column  \textbf{\# Sub$_{m}$} shows the number of submissions that have multiple fixes, and column \textbf{\% Sub$_{m}$} shows what percentage of submissions \textbf{\# Sub$_{m}$} is. 

Overall, only 10.86\% of submissions can be corrected by the current mutant operators. The rate is notably lowest for RL submissions, in which mutants can only correct 5.61\%. In contrast, LTL submissions have nearly quadrupled the fix rate at 19.74\%. This trend holds for submissions with multiple fixes. %, where 1.83\% of RL submissions have multiple corrections but 5.81\% of LTL submissions have multiple corrections. 
Based on classification, overconstrained faults have the most fixes possible at 18.34\% followed by underconstrained faults at 15.48\%, while faults that have both can only be fixed 6.24\% of the time. This is expected, as both over- and underconstrained faults mean the formula is sometimes too permissive and sometimes too restrictive, which is likely to lead to multi-step edits not captured by first-order mutants.

The overall low fix rate is in line with observations from RQ6 in which users over make very similar logical mistakes but the fix is actually two edit steps away, not one. For instance, there is no first order mutation operator that transforms ``\CodeIn{a in b},'' into ``\CodeIn{some a \& b}.'' In addition, there is no first-order mutant operator to transform ``\CodeIn{all a : A | some a.b}'' into ``\CodeIn{some A.b}.'' While second-order mutants would have scalability issues, the common mistakes from RQ6 could be utilized to produce a select subset that reflects common mistakes. Interestingly, mutation operators can correct some type errors, such as transforming ``\CodeIn{s.enrolled.grades}'' from RQ4 into ``\CodeIn{some s.enrolled.grades},'' which could be utilized to improve the Analyzer's current type-based error reports with suggested fixes. 

%Overconstrained faults mean the user was too restrictive, in which case are often addressed by swapping to or inserting a more permissive operator or deleting an operator. In contrast, underconstrained faults mean the formula is too permissive, which often require swapping to  a more restrictive operator or inserting additional constraints to fix, the later of which limits the fixes first-order mutants can provide. Meanwhile, 

\begin{impact}
The current mutant operators do not reflect most mistakes in large part because the most common type of faults, both over- and underconstrained, are rarely mimicked by first order mutants and often require second-order mutants. 
%The most notable gap in mutant effectiveness is also the most common type of fault: both over- and underconstrained errors, which likely require second-order mutants. 
%With only 10% of the faulty submissions fixable, current mutant operators do not reflect most mistakes.  To better capture mistakes that lead to both over- and underconstrained formulas, second-order mutants should be considered.
\end{impact}

\begin{table}
\centering
\caption{Breakdown of Mutant Operators That Fixed Subs}
\label{tab:mutant_op_fix_rate}
\begin{footnotesize}
\begin{tabular}{lrr} \hline
\rowcolor[HTML]{C0C0C0} 
\multicolumn{1}{c}{\cellcolor[HTML]{C0C0C0}\textbf{Mutant Op}} & \multicolumn{1}{c}{\cellcolor[HTML]{C0C0C0}\textbf{\# Fixes}} & \multicolumn{1}{c}{\cellcolor[HTML]{C0C0C0}\textbf{Percent}} \\ \hline
\textbf{Unary Operator Insertion (UOI)} & 1269 & 31.16 \\
\rowcolor[HTML]{EFEFEF} 
\textbf{Unary Operator Deletion (UOD)} & 160 & 3.93 \\
\textbf{Unary Operator Replacement (UOR)} & 340 & 8.35 \\
\rowcolor[HTML]{EFEFEF} 
\textbf{Binary Operator Deletion (BOD)} & 272 & 6.68 \\
\textbf{Binary Operator Replacement (BOR)} & 810 & 19.89 \\
\rowcolor[HTML]{EFEFEF} 
\textbf{Binary Operator Exchange (BOE)} & 167 & 4.10 \\
\textbf{List Operator Deletion (LOD)} & 221 & 5.43 \\
\rowcolor[HTML]{EFEFEF} 
\textbf{List Operator Replacement (LOR)} & 53 & 1.30 \\
\textbf{Quantifier Operator Replacement (QOR)} & 724 & 17.78 \\
\rowcolor[HTML]{EFEFEF} 
\textbf{Prime Operator Insertion (POI)} & 55 & 1.35 \\
\textbf{Predicate Body Deletion (PBD)} & 1 & 0.02 \\ \hline
\end{tabular}
\end{footnotesize}
\end{table}

For viewpoint two, Table~\ref{tab:mutant_op_fix_rate} displays the rate at which individual mutant operators fixed submissions. Column \textbf{\#Fixes} is the number of faulty submissions fixed and \textbf{Percentage} is what percentage of fixed submissions the preceding \textbf{\#Fixes} represents. A majority of the fixes come from replacement (47.32\%) and insertions (32.51\%). 

The high rate of replacement fixes is supported by observations from our common mistakes (RQ6), where we found novice users incorrectly select between similar but different operators. In particular, BOR has the second highest fix rate (19.89\%), which is reflective of the fact that a vast majority of the operators that were mixed up are binary operators. QOR fixes having the third highest fix rate (17.78\%) is also reflective of both choosing the wrong operator, as well as the high prevalence of FOL mistakes. Interestingly, we found that several of the QOR fixes are related to replacing one of the nested quantifiers, such as replacing ``\CodeIn{all a, some b}'' with ``\CodeIn{all a, one b}''. For insertions, UOI fixes were dominant at 31.16\% of all fixes. UOI fixes for LTL operators almost always involve inserting \CodeIn{always}, which is not unexpected as RQ4 highlights that novice users tend to push temporal operators to the front of formulas. For RL, PL, and FOL submissions, the UOI fixes were often the result of inserting transitive closure or transpose, the latter of which is interesting as it means the user was close to the right expression but had inverted part of the relational logic. %This is expected, as we observed common trends where novice users either omitted temporal operators or tended to only place them in the beginning of formulas. 
%For other types of logic, a common UOI correction was inserting the transpose and transitive closure operators. 

\begin{impact}
\vspace{-0.5ex}
Mutant operators that aligned with common mistakes are correlated with the ability to fix more faulty submissions. Therefore, it is worth investing in refining existing and creating new mutant operators based the patterns in RQ6.
\vspace{-0.5ex}
\end{impact}

\subsection{RQ8: How do novice users respond to faulty or invalid submissions?}

As users incrementally work towards a correct solution, users make changes that may move their submission from one classification to the next. These trends can provide insight into how users respond to discovering their submission is faulty or invalid. To explore this, we collected all back-to-back single edit steps for an attempt. Table~\ref{tab:b2b} displays the classification change, or lack thereof, produced by all of these edits. The letter \CodeIn{E} represents empty submissions. Row \textbf{\# Edit `?'} represents the number of edits that start as one classification and move to a new one based on the corresponding column, while Row \textbf{\% Edit `?'} represents what percentage this number is out of the total edits that start as classification `\CodeIn{?}'. For instance, there are 4,133 edits in which the user starts with an overconstrained formula and produces an overconstrained formula (\CodeIn{O-O}). %, which accounts for 40.5\% of the edits that start with an overconstrained formula.

Based on Table~\ref{tab:b2b}, a user is significantly more likely to make an edit that does not change the classification, which implies that users often make small edits. In addition, users do frequently go from an overconstrained to a correct formula (22.92\%), which is in line with our ``one step away'' results in which overconstrained formulas were often close to a correct submission. In contrast, syntax errors, type errors and both submissions are the most likely to lead to a student effectively resetting their attempt by erasing down to an empty submission, which indicates for these categories users are more likely to feel as if they ended up far off from the solution. Interestingly, users do go from a correct submission to all the other classifications. There are two main reasons for this. First, some of the exercises ask for an answer using a specific type of logic, but Alloy4Fun does not restrict the available operators. Therefore, users will create a correct submission and attempt to re-write it into the required logic. Second, users refine their submission, often trying to create a more condense formula. % or produce more readable formulas.

\section{Threats to Validity}
There exist several threats to the validity of our results. First, the dataset we use does not contain any information related to what novice users were thinking. Therefore, we are limited in some of the conclusions we can form. 
%For instance, we can conclude that users give up after 5.85 attempts, but we cannot know if the users gave up because they were frustrated or bored. 
Second, the novice users are all master's students from the University of Minho (UM) and the University of Porto (UP) between the fall of 2019 and the spring of 2023 semesters. While there is breadth to the date range of the study, we do not know the details of the educational performance and background of the master's students, which may lack diversity. Third, MuAlloy, and the Analyzer itself, do not support higher-order quantification. Occasionally, mutating formulas produces higher-order formulas that can not be solved with the current tools. In theory, these higher order formula could be correct. However, since the current tools cannot evaluate them, we do not factor higher order formulas into our calculations of how many submissions are one step away. %When presenting the number of submissions that can be fixed by mutations, we include all submissions in this calculation, even if MuAlloy is unable to produce any valid mutants.  %To limit the impact of this issue, 
 %This allows for our percentage of submissions that can be fixed to most accurately reflect the possibilities of the current tool support.

\begin{table}
\centering
\vspace{-2ex}
\caption{Details of Back-to-Back Edits}
\begin{footnotesize}
\begin{tabular}{l|rrrrrrr} \hline
\rowcolor[HTML]{C0C0C0} 
\multicolumn{1}{c|}{\cellcolor[HTML]{C0C0C0}\textbf{Change}} & \multicolumn{1}{c}{\cellcolor[HTML]{C0C0C0}\textbf{O-O}} & \multicolumn{1}{c}{\cellcolor[HTML]{C0C0C0}\textbf{O-U}} & \multicolumn{1}{c}{\cellcolor[HTML]{C0C0C0}\textbf{O-B}} & \multicolumn{1}{c}{\cellcolor[HTML]{C0C0C0}\textbf{O-S}} & \multicolumn{1}{c}{\cellcolor[HTML]{C0C0C0}\textbf{O-T}} & \multicolumn{1}{c}{\cellcolor[HTML]{C0C0C0}\textbf{O-C}} & \multicolumn{1}{c}{\cellcolor[HTML]{C0C0C0}\textbf{O-E}} \\ \hline
\cellcolor[HTML]{C0C0C0}\textbf{\# Edits O} & \multicolumn{1}{r}{4133} & \multicolumn{1}{r}{526} & \multicolumn{1}{r}{1484} & \multicolumn{1}{r}{762} & \multicolumn{1}{r}{877} & \multicolumn{1}{r}{2339} & \multicolumn{1}{r}{84} \\
\rowcolor[HTML]{EFEFEF} 
\cellcolor[HTML]{C0C0C0}\textbf{\% Edits O} & 40.50 & 5.15 & 14.54 & 7.47 & 8.59 & 22.92 & 0.82 \\ \hline
\rowcolor[HTML]{C0C0C0} 
\multicolumn{1}{c|}{\cellcolor[HTML]{C0C0C0}\textbf{Change}} & \multicolumn{1}{c}{\cellcolor[HTML]{C0C0C0}\textbf{U-O}} & \multicolumn{1}{c}{\cellcolor[HTML]{C0C0C0}\textbf{U-U}} & \multicolumn{1}{c}{\cellcolor[HTML]{C0C0C0}\textbf{U-B}} & \multicolumn{1}{c}{\cellcolor[HTML]{C0C0C0}\textbf{U-S}} & \multicolumn{1}{c}{\cellcolor[HTML]{C0C0C0}\textbf{U-T}} & \multicolumn{1}{c}{\cellcolor[HTML]{C0C0C0}\textbf{U-C}} & \multicolumn{1}{c}{\cellcolor[HTML]{C0C0C0}\textbf{U-E}} \\ \hline
\cellcolor[HTML]{C0C0C0}\textbf{\# Edits U} & \multicolumn{1}{r}{548} & \multicolumn{1}{r}{9472} & \multicolumn{1}{r}{2884} & \multicolumn{1}{r}{1464} & \multicolumn{1}{r}{1248} & \multicolumn{1}{r}{2603} & \multicolumn{1}{r}{110} \\
\rowcolor[HTML]{EFEFEF} 
\cellcolor[HTML]{C0C0C0}\textbf{\% Edits U} & 2.99 & 51.68 & 15.73 & 7.99 & 6.81 & 14.20 & 0.60 \\ \hline
\rowcolor[HTML]{C0C0C0} 
\multicolumn{1}{c|}{\cellcolor[HTML]{C0C0C0}\textbf{Change}}& \multicolumn{1}{c}{\cellcolor[HTML]{C0C0C0}\textbf{B-O}} & \multicolumn{1}{c}{\cellcolor[HTML]{C0C0C0}\textbf{B-U}} & \multicolumn{1}{c}{\cellcolor[HTML]{C0C0C0}\textbf{B-B}} & \multicolumn{1}{c}{\cellcolor[HTML]{C0C0C0}\textbf{B-S}} & \multicolumn{1}{c}{\cellcolor[HTML]{C0C0C0}\textbf{B-T}} & \multicolumn{1}{c}{\cellcolor[HTML]{C0C0C0}\textbf{B-C}} & \multicolumn{1}{c}{\cellcolor[HTML]{C0C0C0}\textbf{B-E}} \\ \hline
\cellcolor[HTML]{C0C0C0}\textbf{\# Edits B} & \multicolumn{1}{r}{1918} & \multicolumn{1}{r}{3952} & \multicolumn{1}{r}{19261} & \multicolumn{1}{r}{2871} & \multicolumn{1}{r}{2545} & \multicolumn{1}{r}{2855} & \multicolumn{1}{r}{178} \\
\rowcolor[HTML]{EFEFEF} 
\cellcolor[HTML]{C0C0C0}\textbf{\% Edits B} & 5.71 & 11.77 & 57.36 & 8.55 & 7.58 & 8.50 & 0.53 \\ \hline
\rowcolor[HTML]{C0C0C0} 
\multicolumn{1}{c|}{\cellcolor[HTML]{C0C0C0}\textbf{Change}} & \multicolumn{1}{c}{\cellcolor[HTML]{C0C0C0}\textbf{S-O}} & \multicolumn{1}{c}{\cellcolor[HTML]{C0C0C0}\textbf{S-U}} & \multicolumn{1}{c}{\cellcolor[HTML]{C0C0C0}\textbf{S-B}} & \multicolumn{1}{c}{\cellcolor[HTML]{C0C0C0}\textbf{S-S}} & \multicolumn{1}{c}{\cellcolor[HTML]{C0C0C0}\textbf{S-T}} & \multicolumn{1}{c}{\cellcolor[HTML]{C0C0C0}\textbf{S-C}} & \multicolumn{1}{c}{\cellcolor[HTML]{C0C0C0}\textbf{S-E}} \\ \hline
\cellcolor[HTML]{C0C0C0}\textbf{\# Edits S} & \multicolumn{1}{r}{1214} & \multicolumn{1}{r}{2116} & \multicolumn{1}{r}{3731} & \multicolumn{1}{r}{8371} & \multicolumn{1}{r}{2239} & \multicolumn{1}{r}{2814} & \multicolumn{1}{r}{203} \\
\cellcolor[HTML]{C0C0C0}\textbf{\% Edits S} & 5.87 & 10.23 & 18.03 & 40.46 & 10.82 & 13.60 & 0.98 \\ \hline
\rowcolor[HTML]{C0C0C0} 
\multicolumn{1}{c|}{\cellcolor[HTML]{C0C0C0}\textbf{Change}} & \multicolumn{1}{c}{\cellcolor[HTML]{C0C0C0}\textbf{T-O}} & \multicolumn{1}{c}{\cellcolor[HTML]{C0C0C0}\textbf{T-U}} & \multicolumn{1}{c}{\cellcolor[HTML]{C0C0C0}\textbf{T-B}} & \multicolumn{1}{c}{\cellcolor[HTML]{C0C0C0}\textbf{T-S}} & \multicolumn{1}{c}{\cellcolor[HTML]{C0C0C0}\textbf{T-T}} & \multicolumn{1}{c}{\cellcolor[HTML]{C0C0C0}\textbf{T-C}} & \multicolumn{1}{c}{\cellcolor[HTML]{C0C0C0}\textbf{T-E}} \\ \hline
\cellcolor[HTML]{C0C0C0}\textbf{\# Edits T} & \multicolumn{1}{r}{1116} & \multicolumn{1}{r}{1647} & \multicolumn{1}{r}{3233} & \multicolumn{1}{r}{1720} & \multicolumn{1}{r}{10528} & \multicolumn{1}{r}{2402} & \multicolumn{1}{r}{184} \\
\rowcolor[HTML]{EFEFEF} 
\cellcolor[HTML]{C0C0C0}\textbf{\% Edits T} & 5.36 & 7.91 & 15.52 & 8.26 & 50.54 & 11.53 & 0.88 \\ \hline
\rowcolor[HTML]{C0C0C0} 
\multicolumn{1}{c|}{\cellcolor[HTML]{C0C0C0}\textbf{Change}} & \multicolumn{1}{c}{\cellcolor[HTML]{C0C0C0}\textbf{C-O}} & \multicolumn{1}{c}{\cellcolor[HTML]{C0C0C0}\textbf{C-U}} & \multicolumn{1}{c}{\cellcolor[HTML]{C0C0C0}\textbf{C-B}} & \multicolumn{1}{c}{\cellcolor[HTML]{C0C0C0}\textbf{C-S}} & \multicolumn{1}{c}{\cellcolor[HTML]{C0C0C0}\textbf{C-T}} & \multicolumn{1}{c}{\cellcolor[HTML]{C0C0C0}\textbf{C-C}} & \multicolumn{1}{c}{\cellcolor[HTML]{C0C0C0}\textbf{C-E}} \\ \hline
\cellcolor[HTML]{C0C0C0}\textbf{\# Edits C} & \multicolumn{1}{r}{692} & \multicolumn{1}{r}{752} & \multicolumn{1}{r}{676} & \multicolumn{1}{r}{526} & \multicolumn{1}{r}{659} & \multicolumn{1}{r}{11395} & \multicolumn{1}{r}{51} \\
\rowcolor[HTML]{EFEFEF} 
\cellcolor[HTML]{C0C0C0}\textbf{\% Edits C} & 4.69 & 5.10 & 4.58 & 3.57 & 4.47 & 77.25 & 0.35 \\ \hline
\rowcolor[HTML]{C0C0C0} 
\multicolumn{1}{c|}{\cellcolor[HTML]{C0C0C0}\textbf{Change}} & \multicolumn{1}{c}{\cellcolor[HTML]{C0C0C0}\textbf{E-O}} & \multicolumn{1}{c}{\cellcolor[HTML]{C0C0C0}\textbf{E-U}} & \multicolumn{1}{c}{\cellcolor[HTML]{C0C0C0}\textbf{E-B}} & \multicolumn{1}{c}{\cellcolor[HTML]{C0C0C0}\textbf{E-S}} & \multicolumn{1}{c}{\cellcolor[HTML]{C0C0C0}\textbf{E-T}} & \multicolumn{1}{c}{\cellcolor[HTML]{C0C0C0}\textbf{E-C}} & \multicolumn{1}{c}{\cellcolor[HTML]{C0C0C0}\textbf{E-E}} \\ \hline
\cellcolor[HTML]{C0C0C0}\textbf{\# Edits E} & \multicolumn{1}{r}{111} & \multicolumn{1}{r}{181} & \multicolumn{1}{r}{225} & \multicolumn{1}{r}{261} & \multicolumn{1}{r}{202} & \multicolumn{1}{r}{344} & \multicolumn{1}{r}{581} \\
\rowcolor[HTML]{EFEFEF} 
\cellcolor[HTML]{C0C0C0}\textbf{\% Edits E} & 5.83 & 9.50 & 11.81 & 13.70 & 10.60 & 18.06 & 30.50 \\ \hline
%\hline
%\rowcolor[HTML]{C0C0C0} 
%\multicolumn{1}{c|}{\cellcolor[HTML]{C0C0C0}\textbf{Change}} & \multicolumn{1}{c}{\cellcolor[HTML]{C0C0C0}\textbf{-> O}} & \multicolumn{1}{c}{\cellcolor[HTML]{C0C0C0}\textbf{-> U}} & \multicolumn{1}{c}{\cellcolor[HTML]{C0C0C0}\textbf{-> B}} & \multicolumn{1}{c}{\cellcolor[HTML]{C0C0C0}\textbf{-> S}} & \multicolumn{1}{c}{\cellcolor[HTML]{C0C0C0}\textbf{-> T}} & \multicolumn{1}{c}{\cellcolor[HTML]{C0C0C0}\textbf{-> C}} & \multicolumn{1}{c}{\cellcolor[HTML]{C0C0C0}\textbf{-> E}} \\ \hline
%\cellcolor[HTML]{C0C0C0}\textbf{\# Edits} & \multicolumn{1}{r}{9732} & \multicolumn{1}{r}{18646} & \multicolumn{1}{r}{31494} & \multicolumn{1}{r}{15975} & \multicolumn{1}{r}{18298} & \multicolumn{1}{r}{24752} & \multicolumn{1}{r}{1391} \\
%\rowcolor[HTML]{EFEFEF} 
%\cellcolor[HTML]{C0C0C0}\textbf{\% Edits} & 8.09 & 15.50 & 26.18 & 13.28 & 15.21 & 20.58 & 1.16
\end{tabular}
\end{footnotesize}
\label{tab:b2b}
\end{table}
\section{Related Work}\label{sec:relatedwork}

\textbf{Alloy4Fun.}
The creators of the Alloy4Fun dataset have published an experience paper outlining the first semester that they used Alloy4Fun in their classroom~\cite{alloy4fun}. This paper focuses largely on lessons learned utilizing Alloy4Fun in the classroom but does contain some preliminary analysis of the data collected, such as the rate of correct versus incorrect submissions and commonly reported error messages. This data is from an early benchmark and consists of 9 models and only 5000 executions. Our analysis is more in-depth and spans the significantly larger version of the dataset that contains 17 models and over 97,000 executions.

\textbf{User Studies and Empirical Studies Over Alloy.}
There have been a few user studies exploring how developers work with Alloy~\cite{alloyformalise23study,negpos,nelsonuserstudy}. The most recent study explored the debugging behavior of novice and expert users and discovered that users struggle to refine Alloy predicates using only the visual representation of scenarios~\cite{alloyformalise23study}. Another recent study on novice Alloy users found that if users are shown a small collection of valid and invalid scenarios before writing a predicate, then users better understand what behavior the user should be trying to codify into a constraint~\cite{negpos}.  In addition, there was a user study that explored how users interact with different enumeration strategies~\cite{nelsonuserstudy}. 
%All of these user studies enrich the same tools and education efforts our study promotes. 
While our efforts focus on how to better teach and debug logical constraints, these user studies highlight best practices for how to present and guide users through the output of these constraints.

Tangent to our work, there is a static profile of 1,652 publicly available Alloy models pulled from GitHub that investigates how often users engage with different parts of Alloy's grammar and explores the average size and complexity of Alloy models. This study is not concerned with the accuracy of the constraints within the model but does give suggestions for education based on frequently used and underutilized features. These recommendations are complementary to our recommendations.

\section{Conclusion}\label{sec:conclusion}
Alloy is a commonly used modeling language that is able to represent static and dynamic behavior of a system. Given Alloy's popularity and well supported toolset, there is a growing body of work to debug faulty models and to use the Alloy in an educational setting. To help guide future research directions for both of these avenues, this paper explores a dataset of over 97,000 submissions made by novice users. We highlight several key findings, including common patterns in correct and incorrect novice submissions, how long novice users will attempt an exercise, and the realistic effectiveness of existing mutant operators. %Based on our observations, we provide a series of recommendations for future researchers. 

\bibliographystyle{splncs04}
\bibliography{bib}

\end{document}